\documentclass{aa}  

\usepackage{amsmath}
\usepackage{hyperref}
\hypersetup{
  colorlinks=true,
  allcolors=[rgb]{0,0,0.8},
  pdftitle={The SRG/eROSITA All-Sky Survey: Constraints on AGN feedback in galaxy groups},
  pdfauthor={Bahar et al.},
  }
\usepackage{graphicx} 
\usepackage{lscape}
\usepackage{mathtools}
\usepackage{txfonts}
\usepackage{booktabs}
\usepackage{ulem}
\usepackage{caption}
\usepackage{subcaption}
\usepackage{threeparttable}
\usepackage{tabularx}
    \newcolumntype{L}{>{\raggedright\arraybackslash}X}

\newcommand{\rosi}{eROSITA\xspace}
\newcommand{\xmm}{\textit{XMM-Newton}\xspace}
\newcommand{\chandra}{\textit{Chandra}\xspace}
\newcommand{\rosat}{ROSAT\xspace}
\newcommand{\suzaku}{\textit{Suzaku}\xspace}

\newcommand{\eromapper}{\texttt{eROMaPPer}\xspace}

\newcommand{\erass}{eRASS1\xspace}

\makeatletter
\newcommand{\specialcell}[1]{\ifmeasuring@#1\else\omit$\displaystyle#1$\ignorespaces\fi}
\newcommand{\pushright}[1]{\ifmeasuring@#1\else\omit\hfill$\displaystyle#1$\fi\ignorespaces}
\newcommand{\pushleft}[1]{\ifmeasuring@#1\else\omit$\displaystyle#1$\hfill\fi\ignorespaces}
\makeatother

\usepackage{natbib,twoopt}

\makeatletter
\newcommandtwoopt{\citeads}[3][][]{\href{http://adsabs.harvard.edu/abs/#3}%
{\def\hyper@linkstart##1##2{}%
 \let\hyper@linkend\@empty\citealp[#1][#2]{#3}}}
\newcommandtwoopt{\citepads}[3][][]{\href{http://adsabs.harvard.edu/abs/#3}%
{\def\hyper@linkstart##1##2{}%
 \let\hyper@linkend\@empty\citep[#1][#2]{#3}}}
\newcommandtwoopt{\citetads}[3][][]{\href{http://adsabs.harvard.edu/abs/#3}%
{\def\hyper@linkstart##1##2{}%
 \let\hyper@linkend\@empty\citet[#1][#2]{#3}}}
\newcommandtwoopt{\citeyearads}[3][][]%
{\href{http://adsabs.harvard.edu/abs/#3}
{\def\hyper@linkstart##1##2{}%
 \let\hyper@linkend\@empty\citeyear[#1][#2]{#3}}}
\makeatother

\begin{document} 

\title{The SRG/eROSITA All-Sky Survey}
\subtitle{Constraints on AGN feedback in galaxy groups}

\author{Y.~E.~Bahar\inst{1}\thanks{e-mail: \href{mailto:ebahar@mpe.mpg.de}{\tt ebahar@mpe.mpg.de}},
E.~Bulbul\inst{1},
V.~Ghirardini\inst{1},
J.~S.~Sanders\inst{1},
X.~Zhang\inst{1},
A.~Liu\inst{1},
N.~Clerc\inst{2},
E.~Artis\inst{1},
F.~Balzer\inst{1},
V.~Biffi\inst{3},
S.~Bose\inst{4},
J.~Comparat\inst{1},
K.~Dolag\inst{5,6},
C.~Garrel\inst{1},
B.~Hadzhiyska\inst{7},
C.~Hern\'andez-Aguayo\inst{6},
L.~Hernquist\inst{8},
M.~Kluge\inst{1},
S.~Krippendorf\inst{5,9},
A.~Merloni\inst{1},
K.~Nandra\inst{1},
R.~Pakmor\inst{6},
P.~Popesso\inst{10},
M.~Ramos-Ceja\inst{1},
R.~Seppi\inst{1},
V.~Springel\inst{6}, 
J.~Weller\inst{1,5},
S.~Zelmer\inst{1}
}

\institute{
\inst{1}{Max Planck Institute for Extraterrestrial Physics, Giessenbachstrasse 1, 85748 Garching, Germany}\\
\inst{2}{IRAP, Université de Toulouse, CNRS, UPS, CNES, Toulouse, France}\\
\inst{3}{INAF --- Osservatorio Astronomico di Trieste, via Tiepolo 11, I-34143 Trieste, Italy}\\
\inst{4}{Institute for Computational Cosmology, Department of Physics, Durham University, South Road, Durham, DH1 3LE, UK}\\
\inst{5}{Universit\"ats-Sternwarte M\"unchen, Fakult\"at f\"ur Physik, Ludwig-Maximilians-Universit\"at, Scheinerstr. 1, 81679 M\"unchen, Germany}\\
\inst{6}{Max Planck Institute for Astrophysics, Karl Schwarzschild Str. 1, Garching, 85741, Germany}\\
\inst{7}{Berkeley Center for Cosmological Physics, Department of Physics, University of California, Berkeley, CA 94720, USA}\\
\inst{8}{Center for Astrophysics | Harvard \& Smithsonian, 60 Garden Street, Cambridge, MA 02138, USA}\\
\inst{9}{Arnold Sommerfeld Center for Theoretical Physics, Ludwig-Maximilians Universität, Theresienstr.~37, 80333 München, Germany}\\
\inst{10}{European Southern Observatory, Karl Schwarzschildstrasse 2, 85748, Garching bei M\"unchen, Germany}\\
}

\titlerunning{Constraints on AGN Feedback in the eROSITA selected galaxy groups}
\authorrunning{Bahar et al.}

\abstract
{Galaxy groups lying between galaxies and galaxy clusters in the mass spectrum of dark matter halos play a crucial role in the evolution and formation of the large-scale structure. Their shallower potential wells compared to clusters of galaxies make them excellent sources to constrain non-gravitational processes such as feedback from the central active galactic nuclei (AGN).
}
{We investigate the impact of feedback, particularly from AGN, on the entropy and characteristic temperature measurements of galaxy groups detected in the SRG/eROSITA's first All-Sky Survey (eRASS1) to shed light on the characteristics of the feedback mechanisms and help guide future AGN feedback implementations in numerical simulations.}
{
We analyzed the deeper \rosi\ observations of 1178 galaxy groups detected in the eRASS1. We divided the sample into 271 subsamples based on their physical and statistical properties and extracted average thermodynamic properties, including the electron number density, temperature, and entropy, at three characteristic radii from cores to outskirts along with the integrated temperature by jointly analyzing X-ray images and spectra following a Bayesian approach.
}
{
We present the tightest constraints with unprecedented statistical precision on the impact of AGN feedback through our average entropy and characteristic temperature measurements of the largest group sample used in X-ray studies, incorporating major systematics in our analysis. We find that entropy shows an increasing trend with temperature in the form of a power-law-like relation at the higher intra-group medium (IGrM) temperatures, while for the low-mass groups with cooler ($T<1.44$~keV) IGrM temperatures, a slight flattening is observed on the average entropy. Overall, the observed entropy measurements agree well with the earlier measurements in the literature. Additionally, comparisons with the state-of-the-art cosmological hydrodynamic simulations (MillenniumTNG, Magneticum, OWL) after applying the selection function calibrated for our galaxy groups reveal that observed entropy profiles in the cores are below the predictions of simulations. At the mid-region, the entropy measurements agree well with the Magneticum simulations, whereas the predictions of MillenniumTNG and OWL simulations fall below observations. At the outskirts, the overall agreement between the observations and simulations improves, with Magneticum simulations reproducing the observations the best.
}
{These measurements will pave the way for achieving more realistic AGN feedback implementations in numerical simulations. The future eROSITA Surveys will enable the extension of the entropy measurements in even cooler IGrM temperatures below $0.5$~keV, allowing for the testing of the AGN feedback models in this regime.}

\maketitle
%
%
\section{Introduction}

In the current understanding of the "bottom–up" structure formation of the Universe in the standard $\Lambda$CDM cosmology, small overdensities collapse first, overcoming the cosmological expansion and merging to form larger halos \citep{Springel2005}. In this scenario, the gas encapsulated in dark matter halos forms the first stars and galaxies as it cools and condenses. The effects of tidal forces, mergers, and interactions in their surroundings regulate the galaxy formation and evolution process. The majority of galaxies in the Universe are found in dense environments as galaxy groups and include a large fraction of the universal baryon budget \citep{Mulchaey2000}. The interaction between galaxies and the intra-group medium (IGrM; the gas encapsulated within the galaxy groups) plays a crucial role in shaping the properties and evolution of galaxies. For instance, the feedback from supernovae, star formation, and central supermassive black holes impacts the evolution of galaxies within the galaxy groups. 

Although there is no clear definition of galaxy groups, dark matter halos with fewer than 50 galaxies and/or masses between $5\times10^{12}-10^{14}$~M$_{\odot}$ are classified as galaxy groups \citep{Crain2009} in the literature. In addition to their member galaxies, galaxy groups contain diffuse baryonic matter in the form of hot plasma with temperatures ($T$\footnote{Throughout this paper, we use the notation $T$ to represent $k_{B}T$ and express temperature measurements in the units of kiloelectron volt.}) ranging from 0.1--2~keV, all encapsulated by the potential well provided by dark matter. Groups are further categorized as loose (or poor) groups, compact groups, and fossil groups, depending on their optical properties \citep{Hickson1997, Mulchaey1998, Mulchaey2000, Voevodkin2010}. 

Despite the abundance of galaxy groups and the essential role they play in the assembly process of dark matter halos in the Universe \citep{Crain2009}, their detection has been challenging due to their low richness, faint X-ray signal, and shallow potential wells. A variety of methods have been employed to search for galaxy groups. In the optical domain, clustering and friends-of-friends (FoF) algorithms have been used to catalog groups in spectroscopic or photometric galaxy redshift surveys \citep[e.g.,][]{Hickson1982, Robotham2011, Tempel2017, Gozaliasl2022}. However, due to their relatively low richnesses, group catalogs compiled using optical observations may suffer from large contamination fractions from the random superposition of galaxies along the line-of-sight, otherwise known as projection effects \citep[see][]{Costanzi2019, Grandis2021, Myles2021}. On the other hand, in the X-ray domain, the emission from IGrM makes them appear as extended sources in the X-ray sky, where they typically exhibit rapidly increasing X-ray emission profiles from the outskirts of the system to the center. Because of their characteristic surface brightness profiles, if their emission is above the background level, they can be easily identified and do not suffer significantly from projection effects. Aside from being a reliable tool for detecting groups, X-ray observations also enable the measurement of the physical properties of the hot ionized IGrM through imaging and spectral analysis.

The effects of the non-thermal astrophysical phenomena governing galaxy formation are easier to study using galaxy groups compared to clusters, as the input energy associated with these phenomena is comparable to the binding energies of groups \citep{Balogh2001}. For investigating the non-thermal phenomena, entropy ($S= T/n_{\rm e}^{2/3}$) measurements of IGrM are often used, which retain a historical record of the thermodynamic state of the gas and reflect the changes in the cooling and heating processes in galaxy groups \citep{Voit2005,Pratt2010}. For example, OverWhelmingly Large Simulations \citep[OWLS,][]{Schaye2010,McCarthy2010} showed that the outflows from the central active galactic nuclei (AGN) elevate the entropy of the IGrM by mechanically expelling low-entropy gas instead of directly heating it. This process mitigates rapid cooling and prevents excessive star formation \citep{Bryan2000, Balogh2001}. On the other hand, the feedback generated by supernova-driven winds associated with galaxies can also increase the entropy of the IGrM through direct heating \citep[see][for a review]{Eckert2021}.

Another important observational constraint on the non-thermal astrophysical phenomenon is the shape of the stellar mass function. Observational studies have demonstrated that the stellar mass exhibits a cutoff at $M_{\star}\sim 10^{11}M_{\odot}$ \citep{Davidzon2017}. Early studies trying to reproduce the cutoff with stellar feedback (e.g., through the energy and momentum released by supernovae explosions and stellar winds) were unsuccessful, as the injected energy was proven to not be enough to prevent cooling and regulate the star formation efficiency \citep{Benson2003}. Therefore, it is commonly agreed that feedback from another source, such as  AGN, is needed to reproduce the observed cutoff in the stellar mass function \citep{Harrison2017}. Moreover, there are further observational constraints on the energy released by the non-gravitational feedback, such as the Si and Fe abundance profiles of the intracluster and intragroup mediums. These measurements cannot be reproduced even with the assumption of 100 percent efficient stellar feedback heating \citep{Kravtsov2000}, a scenario that is rejected by the measurements of the galactic outflows \citep{Martin1999}. Therefore, the total amount of energy that can be injected through stellar feedback on the IGrM is constrained relatively well by the abundant observational data. Consequently, by measuring the thermodynamic properties of galaxy groups, one effectively constrains the energetics of the remaining source of energy, AGN. For the higher mass groups ($M_{500c} > 10^{13.5}~M_{\odot}$), the impact of stellar feedback is at a negligible level such that entropy measurements put direct constraints on the impact of AGN \citep{LeBrun2014}. For the low-mass groups ($M_{500c} < 10^{13}~M_{\odot}$), constraints from multiple observables should be combined to isolate the impact of AGN on its surroundings \citep[e.g., see][]{Altamura2023}.

The AGN heating in galaxy clusters and groups is observationally confirmed by shocks, ripples, and cavities detected in X-ray wavelengths \citep[e.g.,][]{Fabian2006,Randall2011} as well as the detection of radio-loud AGN in a significant proportion of the brightest cluster and group galaxies of the cool core galaxy clusters and groups \citep[e.g.,][]{Burns1990,Best2007,Smol2011}. Furthermore, deeper radio observations have revealed that nearly every central galaxy in X-ray bright groups hosts radio emission \citep{Kolokythas2019}. In fact, radio observations of galaxy groups are highly complementary to the X-ray view of groups for investigating the impact of AGN on IGrM \citep{Eckert2021}. Simultaneously studying their X-ray and radio properties allows for putting constraints on the radio mode feedback from the central engine \citep[e.g.,][]{Pasini2022,Bockmann2023}. Nevertheless, combining multi-wavelength datasets comes with challenges. For instance, crossmatching X-ray and radio catalogs makes it challenging to have a good handle on the selection effects, which is crucial for achieving unbiased conclusions about galaxy groups at the population level. Given the challenges and caveats, in this work, we only focus on putting constraints on the impact of non-gravitational feedback mechanisms through X-ray observations and leave the investigation of the multi-wavelength picture of the eRASS1 galaxy groups sample to future work.

Entropy of IGrM can be measured using X-ray observations, where the electron density and temperature measurements can be made using the imaging and spectroscopic capabilities of X-ray telescopes, such as SRG/eROSITA, \xmm, and {\it Chandra}. \citet{Ponman1999} measured the entropy of 25 bright galaxy clusters and groups at a radius of $0.1r_{virial}$\footnote{The term $r_{virial}$ is defined as the radius within which a system obeys the virial theorem.} using \rosat and GINGA observations and reported that the entropy measurements at the core lie above the expected power-law relation with temperature for the first time. Subsequently, \citet{Lloyd-Davies2000}, \citet{Finoguenov2002}, and \citet{Ponman2003} measured the entropy profiles of galaxy clusters and groups using \rosat and ASCA observations that provided the first hint that in galaxy groups, the excess entropy is not limited only to the core but can also be prominent at larger radii. \citet{Voit2005} formulated a baseline entropy profile that can be used for evaluating the impact of non-gravitational processes for galaxy clusters and groups using four sets of simulations that only include gravitational processes. Using X-ray instruments with a higher spatial and spectral resolution, such as \xmm and \chandra, significantly improved our understanding of the excess entropy in galaxy groups by accurately measuring their entropy profiles. \citet{Johnson2009} investigated entropy profiles of galaxy groups by analyzing \xmm observations of 28 nearby galaxy groups from the Two-Dimensional \xmm Group Survey. They divided their sample into two subsamples (cool core and non-cool core) based on the temperature gradient at the core of their groups and found that the entropy profiles of the groups in their non-cool core sample exhibit less scatter compared to entropy profiles of their cool core sample. Around the same time, \citet[S09 hereafter]{Sun2009} conducted a comprehensive study on the thermodynamic gas properties of 43 nearby galaxy groups using the archival \chandra observations, where they constrained the temperature, electron density, and metallicity profiles of 23 groups out to $r_{500c}$\footnote{The term $r_{500c}$ is defined as the radius within which the density of a system is 500 times the critical density of the Universe at the redshift of the system.} accurately thanks to the outstanding imaging capabilities of \chandra and the relatively deep archival observations of some systems in their sample. Furthermore, they compared their entropy profiles with the baseline entropy profile of \citet{Voit2005} and found that even though the entropy excess reduces as a function of the radius, it remains significant out to $r_{500c}$. Subsequently, the detailed analysis of the outskirts of RX J1159+5531, UGC 03957, and Virgo using \suzaku observations revealed that the entropy excess can go beyond $r_{500c}$ \citep{Humphrey2012,Tholken2016,Simionescu2017}. More recently, \citet{Panagoulia2014} analyzed 66 galaxy groups from the NORAS and REFLEX samples to investigate the properties of IGrM at the core and found that entropy profiles of galaxy groups at the core follow a power-law relation and do not exhibit any entropy floor.

Previous studies in the literature on the thermodynamic properties of galaxy groups have been conducted using relatively small (fewer than 100) and highly incomplete samples that lack well-defined selection functions. Notably, eROSITA opens a new window for galaxy group studies by providing the largest pure X-ray selected sample with a well-defined selection function, which is crucial for achieving robust conclusions that reflect the physical properties of the galaxy group well at the population level. Furthermore, the superb soft X-ray band sensitivity and the scanning observing strategy of eROSITA make it an excellent instrument for investigating the physical properties of the hot gas in galaxy groups, as the emission of IGrM peaks at the soft X-ray band and the brightest galaxy groups above the detection capabilities of the current instruments are at low redshift and well extended.

In this work, we examine the effect of the feedback on the thermodynamics of galaxy groups detected by eROSITA in its first All-Sky Survey. We accomplish this by performing joint imaging and spectral analysis on the eRASS:4 (the four consecutive eROSITA All-Sky Surveys stacked together) observations of the galaxy groups in our sample. The extended ROentgen Survey with an Imaging Telescope Array (\rosi), the soft X-ray telescope on board the Spectrum-Roentgen-Gamma (SRG) mission \citep{Sunyaev2021}, was launched on July 13, 2019 \citep{Predehl2021}. The first All-Sky Survey with eROSITA was successfully executed on June 11, 2020, after 184 days of operation.  In this first All-Sky Survey (eRASS1), eROSITA detected a total of 12247 optically confirmed galaxy groups and clusters spanning the redshift range $0.003 < z < 1.32$ with a sample purity level of 86\% in the Western Galactic half of the survey (359.9442~deg~$> l >$~179.9442~deg), where the data rights belong to the German \rosi\ consortium \citep{Merloni2024, Bulbul2024, Kluge2024}.

In this paper, we combine the imaging and spectroscopic information of 1178 eROSITA-detected galaxy groups and obtain average entropy measurements at $0.15r_{500c}$, $r_{2500c}$ and $r_{500c}$ to investigate the effects of AGN feedback and compare our findings with the state-of-the-art numerical simulations from  MillenniumTNG \citep{Hernandez2023,Pakmor2023}, Magneticum \citep{Hirschmann2014}, and OverWhelmingly Large Simulations \citep{Schaye2010,McCarthy2010}. The findings represent the first study of a comprehensive group sample with a well-defined selection function. This paper is organized as follows: In Sect.~\ref{sec:sample_selection}, we describe the construction of the galaxy groups sample from the primary eRASS1 sample, and in Sect.~\ref{sec:data_analysis}, we describe the X-ray data reduction and the analysis of the groups. In Sect.~\ref{sec:systematics}, we provide a discussion of major systematics and the details of the quantification and incorporation of them in our results, and in Sect.~\ref{sec:results}, we provide our final results on the entropy measurements of the sample and comparisons with the previous measurements. In Sect.~\ref{sec:sim_comparison}, we provide a comparison between our measurements and the predictions of the state-of-the-art simulations. Lastly, we provide a summary of our findings and list our conclusions in Sect.~\ref{sec:conclusions}. Throughout this paper, we adopt a flat $\Lambda$CDM cosmology using the \citet{Planck2016} results, namely $\Omega_m =0.3089$, $\Omega_b =0.0486$, $\sigma_8=0.8147,$ and $H_0 = 67.74$~km~s$^{-1}$~Mpc$^{-1}$. Quoted error bars correspond to a 1-$\sigma$ confidence level unless noted otherwise.

\begin{figure*}
\includegraphics[width=0.60\textwidth]{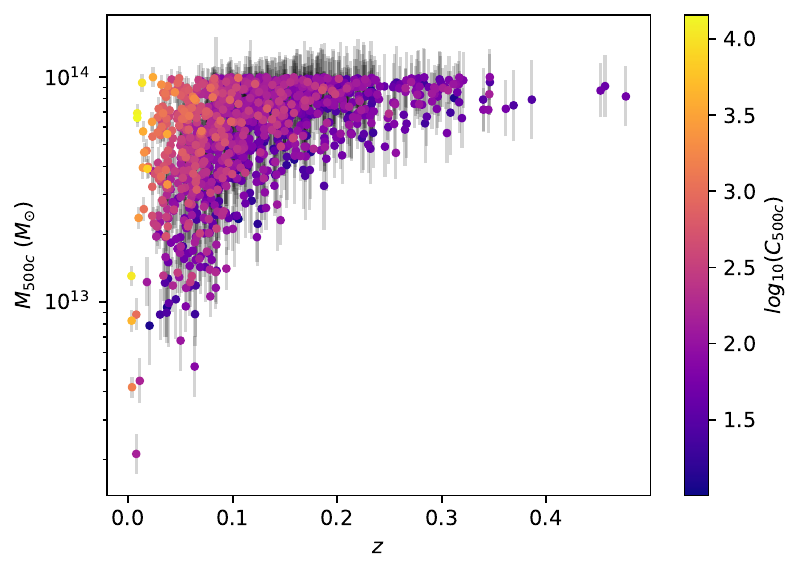} 
\includegraphics[width=0.39\textwidth]{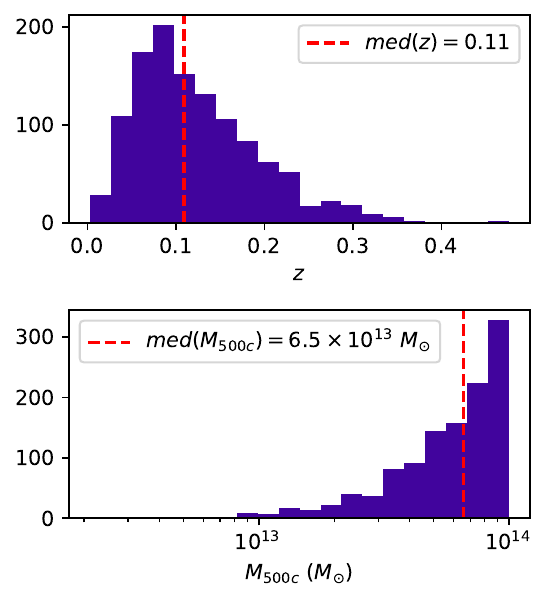} 
\caption{{\it Left:} Mass and redshift distributions of the galaxy group sample used in this work consisting of 1178 objects where colors of the data points represent the eROSITA counts of the groups in soft X-ray band (0.3 -- 1.8~keV) {\it Right:} Mass and redshift histograms of the group sample where the median redshift ($z$) is 0.11 and median mass ($M_{500c}$) is $6.5 \times 10^{13}~M_{\odot}$. \label{fig:mass_z_hist}}
\end{figure*}
\section{Sample of galaxy groups}
\label{sec:sample_selection}

This work utilizes a subsample of the X-ray selected, optically identified primary eRASS1 galaxy cluster and group sample detected in the Western Galactic hemisphere of the first eROSITA All-Sky Survey \citep{Bulbul2024, Kluge2024}. Below, we briefly describe the detection of galaxy clusters and groups in eRASS1 observations and a brief summary of the optical and X-ray cleaning performed in \citet{Bulbul2024}. Subsequently, we provide the details of the additional selection and cleaning applied to the eRASS1 galaxy clusters and groups catalog. 

The X-ray emitting celestial objects in the eRASS1 master X-ray catalog \citep{Merloni2024} are detected using the eROSITA source detection pipeline, which is part of the eROSITA Science Analysis Software System \citep[\texttt{eSASS}\xspace,][]{Brunner2022}. The pipeline locates detection candidates and calculates detection and extent likelihood ($\mathcal{L}_{\rm det}$ and $\mathcal{L}_{\rm ext}$) parameters by comparing the spatial distribution and the abundance of the photons around the candidate with the local background. To construct the primary galaxy groups and clusters sample \citep{Bulbul2024}, a $\mathcal{L}_{\rm ext}>3$ cut is applied to increase the completeness of the galaxy groups and clusters sample \citep[see][for the motivation]{Bulbul2022}. The DESI Legacy Survey DR9 and DR10 datasets are used in the optical identification processes by the \texttt{eROMaPPer} pipeline, which is based on the matched-filter red-sequence algorithm from \texttt{redMaPPer} \citep{Rykoff2014, Rykoff2016} tailored and optimized for the identification of eROSITA extended sources \citep{IderChitham2020, Kluge2024}. If available, spectroscopic redshifts ($z_{\rm spec}$) are prioritized over photometric redshift ($z_{\lambda}$) by the \texttt{eROMaPPer} pipeline \citep[see][for further details]{Kluge2024}. In this primary sample, 12\,705 extended sources in the redshift range of 0.01 to 1.35 are identified as galaxy clusters or groups with a contamination fraction of 14\% \citep{Bulbul2024, Kluge2024}.

To construct a final clean and secure galaxy group sample, we apply further cuts based on the X-ray and optical properties of the primary sample. While the literature lacks a precise definition for galaxy groups, we classify an object as a group if its mass ranges between $5\times10^{12}<M_{500c}< 10^{14} M_{\odot}$. The upper end of our group definition ($10^{14} M_{\odot}$) corresponds to a plasma temperature of $T \sim 2$ keV and is commonly used in previous X-ray studies for distinguishing galaxy clusters from galaxy groups \citep{Lovisari2021}. For incorporating this mass criterion, we use the $M_{500c}$ estimates obtained in Sect.~\ref{sec:ltm_estimation} using a Bayesian X-ray observable estimation framework that jointly estimates the soft-band (0.5--2~keV) X-ray luminosity ($L_{X}$), temperature ($T$) and the mass ($M_{500c}$) of galaxy clusters and groups from their count-rate profiles (see Sect.~\ref{sec:ltm_estimation} for the details of our $L_{X}-T-M_{500c}$ estimation). After applying a mass cut of $M_{500c}<10^{14} M_{\odot}$, we select 2526 galaxy group candidates with a median redshift of $0.11$.

To further reduce the contamination, we apply other cleaning methods using the deeper eRASS:4 data. Contaminants in the $\mathcal{L}_{\rm ext}>3$ sample of the eRASS1 clusters and groups catalog can be classified into two categories: misclassified sources (mostly AGN) and spurious sources. Given that our preliminary sample has a median redshift of 0.11, the "real" galaxy groups in our sample are expected to be relatively well extended in the sky, whereas misclassified point sources, by definition, should have a low extent. We make use of this fact and conservatively remove 841 objects that have ${\tt EXT} < 20$~arcsec and $\mathcal{L}_{\rm ext} < 5.5$. These cuts remove most point sources, leaving 1685 group candidates in the sample.

Once the misclassified point sources are removed, spurious sources are left to be cleaned from our group sample. We use the count measurements in the 0.3 -- 1.8~keV band (see Appendix~\ref{sec:ebandselection} for the details on the choice of the energy band) obtained from eRASS:4 observations as described in Sect.~\ref{sec:imaging_analysis} to clean the spurious sources. We first apply an X-ray count cut of 10. This cut removes 423 objects from our group candidates. Furthermore, we remove 10 more sources from the remaining sample with count measurements $1\sigma$ consistent with the background level. This procedure removes most of the spurious sources since one would expect the galaxy groups to be more prominent and bright as the survey gets deeper. On the other hand, the spurious sources are expected to have low counts and be consistent with the background level since they are mostly due to background fluctuations or superpositions of undetected AGN in the eRASS1 observations. Applying these cuts, we remove a large fraction of contaminants in our sample and obtain a highly pure sample with 1252 galaxy groups. 

One of the major benefits of the strict cleaning procedure described above is its applicability to simulations. Our cleaning procedure relies on the detection pipeline outputs (e.g., $\mathcal{L}_{\rm ext}$ and ${\tt EXT}$), and therefore the selection process is fully reproducible in the simulations of the eRASS1 digital twin \citep{2020Comparat, Seppi2022}. This allows us to construct a robust selection function for our sample using the \rosi's digital twin simulations. We further note that the cleaning applied in this work to remove spurious sources has no impact on the selection function.\footnote{The selection function, $P(I|O)$, can be seen as the ratio of the number of the detected "real" objects to the number of all the "real" objects within the infinitesimal observable parameter space of ($O$,$O+dO$). Removal of confirmed spurious sources does not change this ratio and, therefore, has no impact on $P(I|O)$.}

\begin{figure*}
\begin{centering}
\begin{tabular}{c}
\includegraphics[width=0.98\textwidth,trim={0.1cm 1.5cm 0.1cm 1.5cm}, clip]{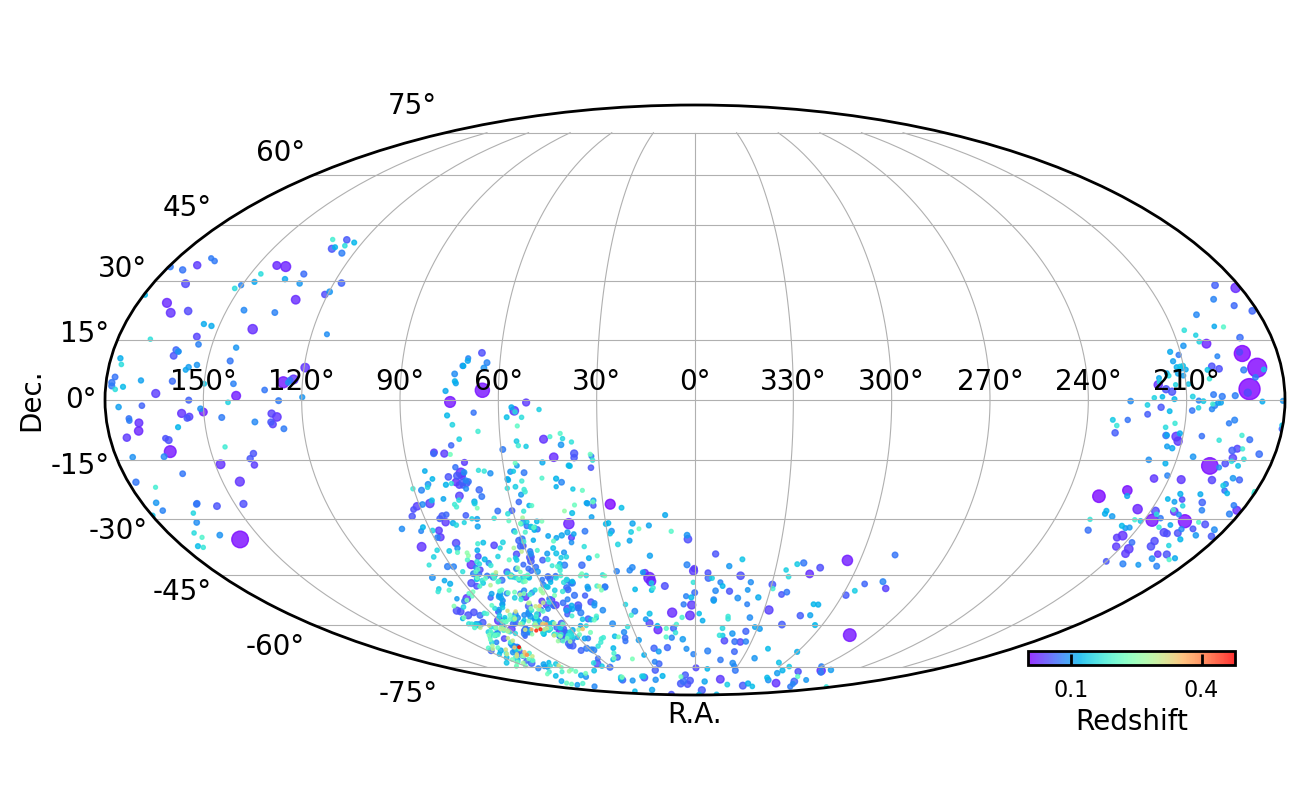} 
\end{tabular}
\end{centering}
\vspace{-0mm}
\caption{Projected locations of the 1178 galaxy groups in the primary catalog in the \rosi\ and Legacy Survey DR9N and DR10 13,116~deg$^2$ common footprints. The redshift confirmed by the follow-up algorithm \eromapper\ is color coded \citep{Kluge2024}, while the sizes of the detections are scaled with the angular sizes ($r_{500c}$) of the groups (see Sect.~\ref{sec:ltm_estimation} for the $r_{500c}$ estimation procedure). The inhomogeneity of the source density in this figure is due to the exposure variation across the eROSITA-DE X-ray sky \citep[see Fig. 2 in ][]{Bulbul2024}.\label{fig:projimage}}
\end{figure*}

Following the cleaning procedure, we visually inspect the groups that are centrally peaked ($r_{c}$ < 32 arcsec)\footnote{See Eq.~\ref{eqn:vikhlinin_ne} for the definition of $r_{c}$ and Sect.~\ref{sec:imaging_analysis} for the details of the imaging analysis.} and have low extent likelihoods ($\mathcal{L}_{ext}<10$) using the eRASS:4 and the Legacy Survey data. As a result of the visual inspection, we further flagged and removed 74 falsely classified point sources from our sample. These objects make a small fraction of our clean sample ($6 \%$); therefore, their removal has a negligible impact on the selection function, especially compared to the systematic uncertainties of X-ray simulations at group scales used to construct the selection function. The final sample, consisting of 1178 galaxy groups, has a median redshift of $0.11$ and a median mass ($M_{500c}$) of $6.3 \times 10^{13}$~M$_{\odot}$. The Redshift and mass distributions of the final sample can be seen in the right panel of Fig.~\ref{fig:mass_z_hist}. Moreover, the 2D projected distribution of these groups in the \rosi\ sky is shown in Fig.~\ref{fig:projimage}. We note that some of the "cleaning" procedures described above (e.g., the ${\tt EXT}$ and $\mathcal{L}_{\rm ext}$ cuts) not only clean spurious sources but also unavoidably remove some of the faint real groups from the sample according to the expected purity of the eRASS1 cluster sample in the cost of achieving a more secure groups sample. Nevertheless, the resulting extra selection is taken into account in our analysis by incorporating a selection function built for our final sample. After the cleaning procedure described above, our sample ended up having three objects with mass estimates slightly below the lower bound of our group mass definition ($5\times 10^{12}~M_{\odot}<$~M~$_{500c}<10^{14}~M_{\odot}$). We eventually decided to keep them in our group sample since the removal of three objects has little to no impact on our final results, and their "true" masses can well be within our group mass definition due to the intrinsic scatter of the $L_{X}-M_{500c}$ relation.

The final galaxy group sample described above is obtained to construct a well-defined selection function using \rosi's twin simulations. A good handle on selection effects is key for achieving universal conclusions about the properties and the governing physics of studies of groups. The deeper eRASS:4 observations of an unprecedented number of galaxy groups we use in this work are particularly well-suited for studying the baryonic physics in galaxy groups because of the higher statistics allowed by the deeper survey data and large field of view necessary to measure the X-ray properties out to large radii. In the next section, we present our eRASS:4 analysis of the galaxy groups in the sample.

\begin{figure*}[h!]
\begin{center}
\includegraphics[width=0.49\textwidth]{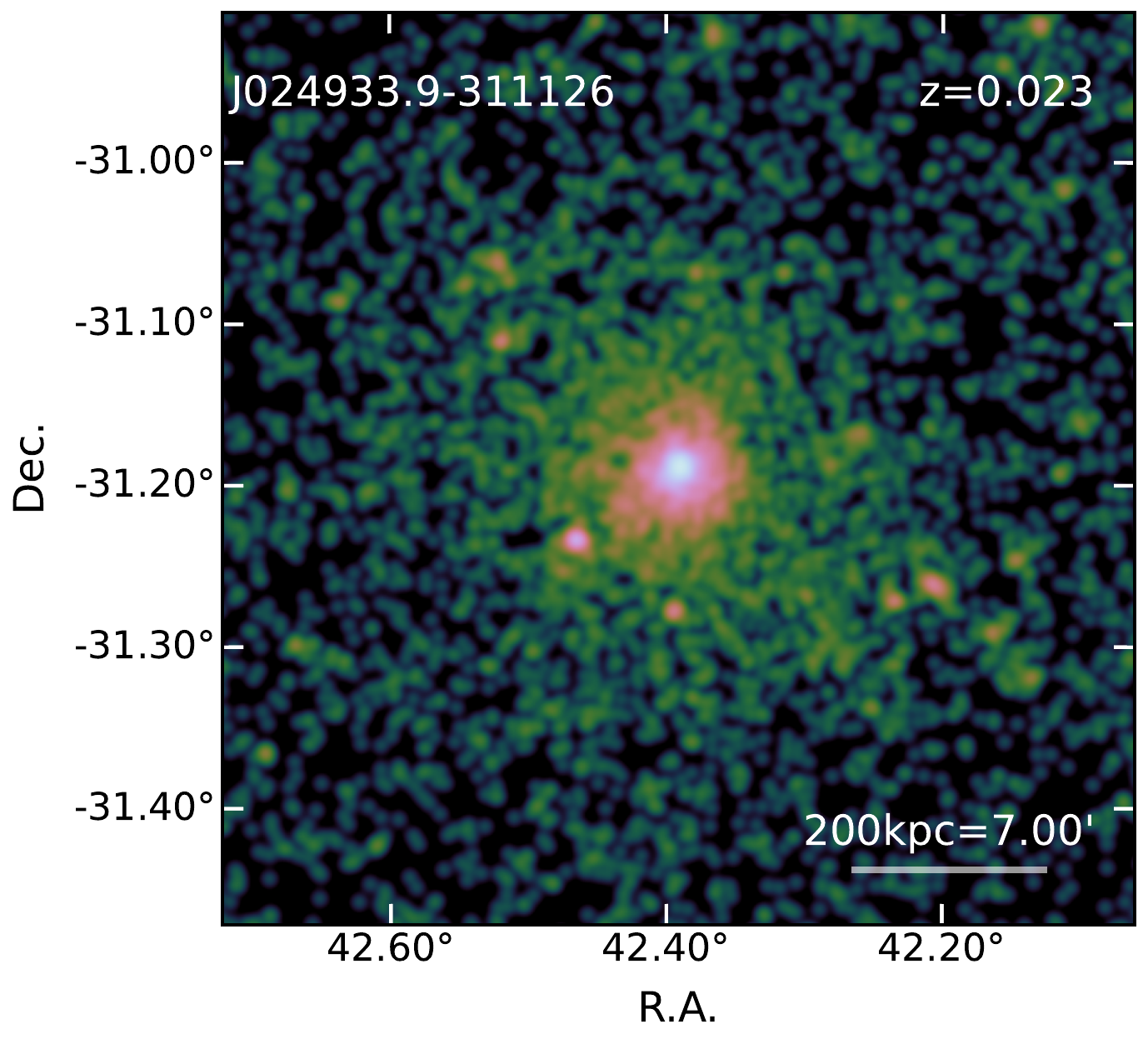} 
\includegraphics[width=0.49\textwidth]{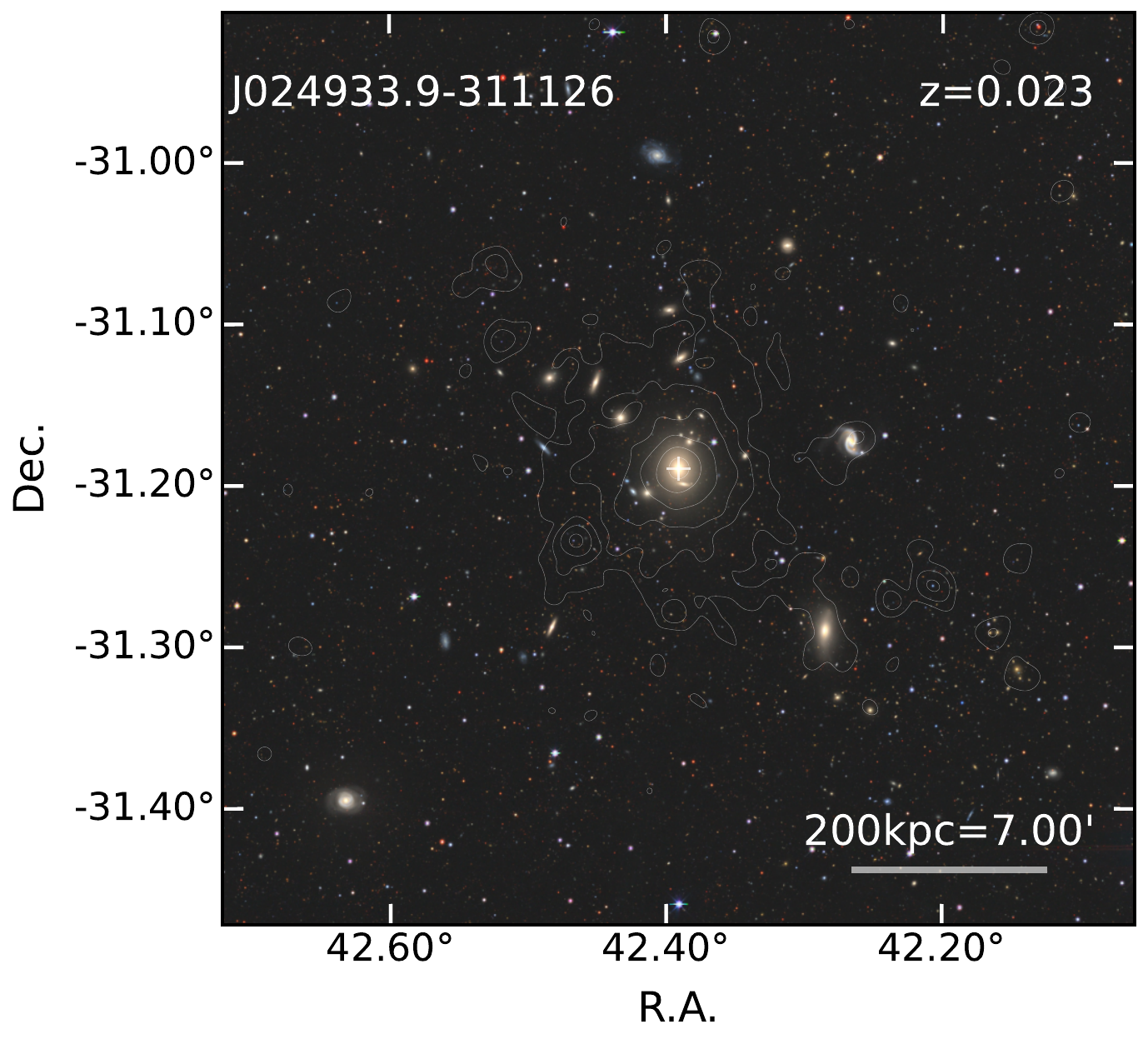} \\
\includegraphics[width=0.49\textwidth]{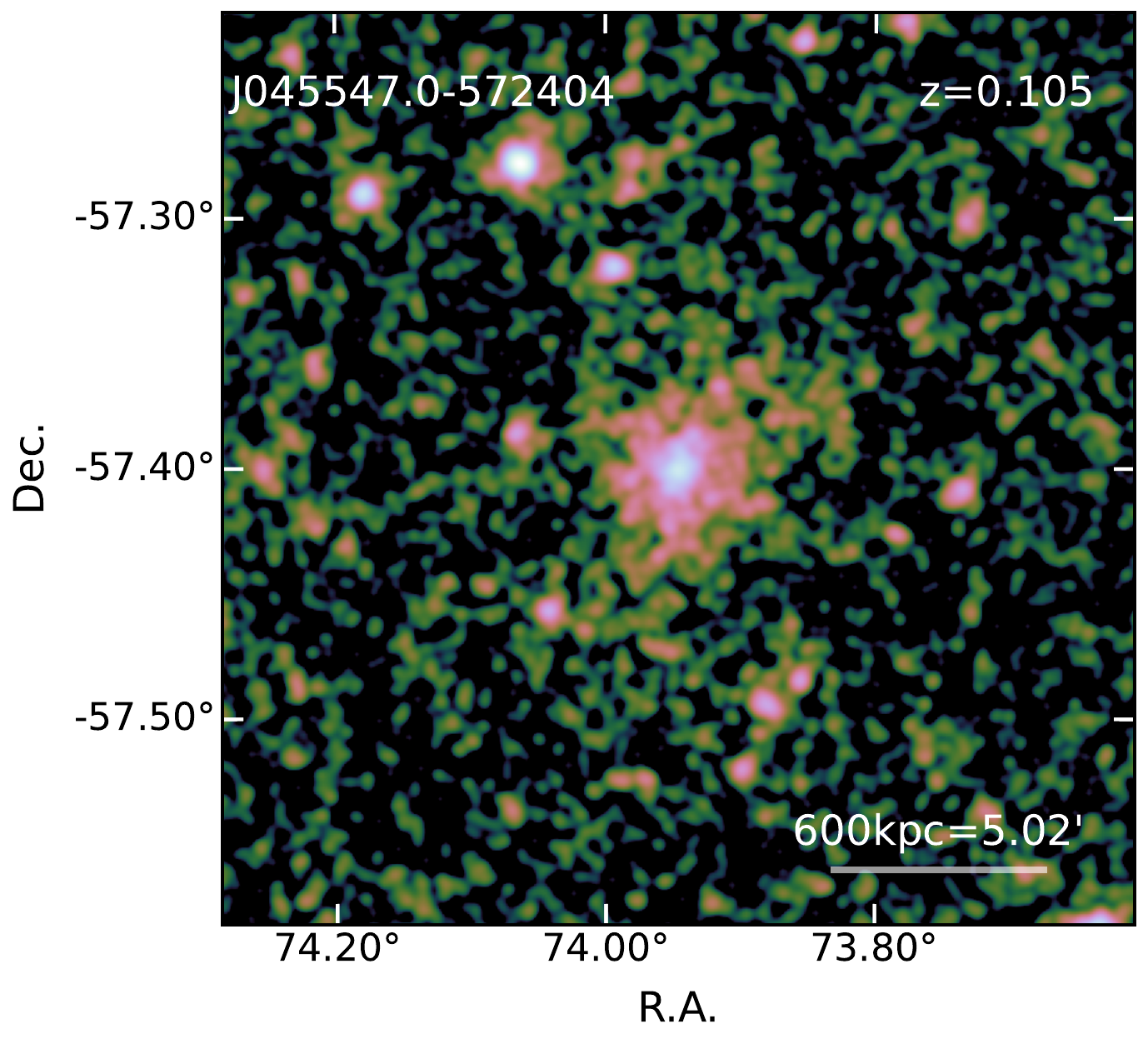} 
\includegraphics[width=0.49\textwidth]{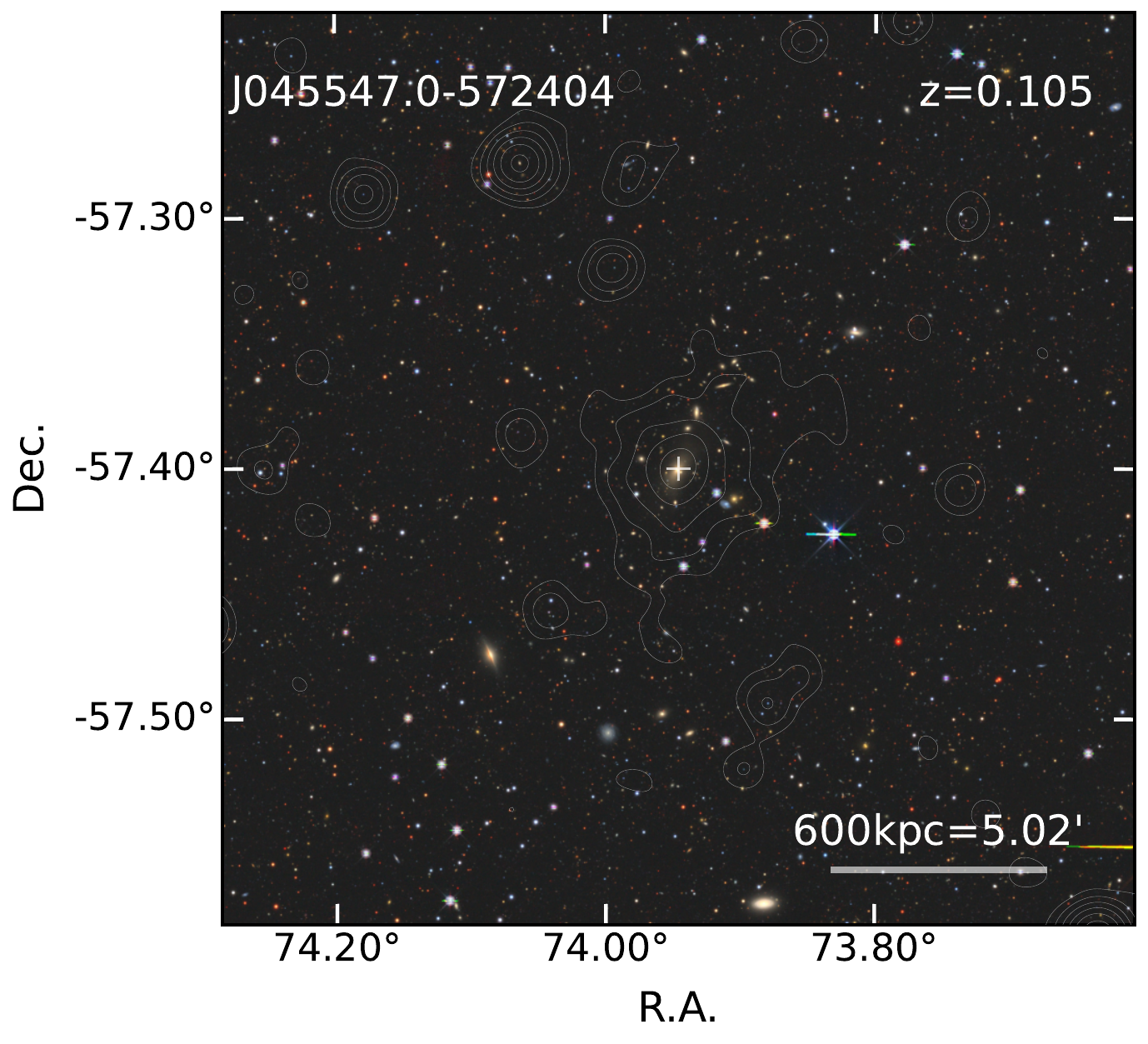} 
\end{center}
\caption{{\it Left:} eRASS:4 soft band 0.3--1.8~keV images of two bright groups (1eRASS~J024933.9-311126 and 1eRASS~J045547.0-572404) at redshifts 0.023 and 0.105. {\it Right:} Legacy Survey DR10 images of the same groups with the eRASS:4 X-ray contours overlayed.\label{fig:xray_optical_example}}
\end{figure*}
\begin{figure*}
\begin{center}
\includegraphics[width=0.99\textwidth,trim={30 0 0 10}, clip]{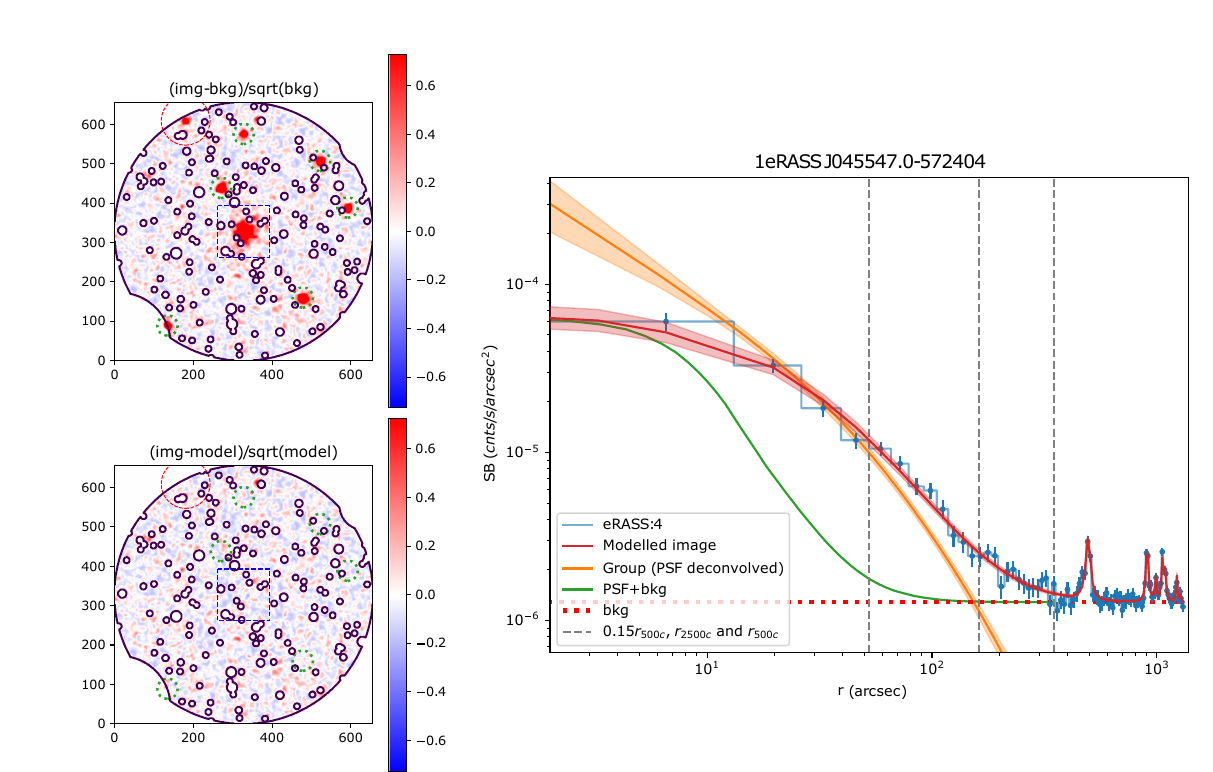} 
\end{center}
\caption{Example of the \rosi\ imaging analysis for the group 1eRASS~J045547.0-572404. {\it Left:} Residual image of the group before and after subtracting the co-fit extended sources and point sources in the field. Nearby co-fit clusters and groups are shown with red circles, co-fit nearby bright AGN are shown with green circles, and the fitted galaxy group is shown with a blue rectangle. The smooth noise level indicates that the contaminant emission is modeled properly in the analysis. {\it Right:} Surface brightness profile of the same group. The observed surface brightness profile of the image is plotted in blue, the best-fit model of the image is plotted in red, the PSF profile over the measured background is plotted in green, the PSF deconvolved surface brightness profile of the galaxy group is plotted in orange, the measured background level is shown with a horizontal dashed red line, and three characteristic radii of the group ($0.15r_{500c}$, $r_{2500c}$ and $r_{500c}$ from the core to the outskirt respectively) are shown with dashed gray lines.\label{fig:residual_crprof}}
\end{figure*}

\section{Data analysis}
\label{sec:data_analysis}

\subsection{X-ray data reduction and analysis}

Taking advantage of the higher signal-to-noise, deeper survey observation, we used the eRASS:4 observations of the eRASS1 selected galaxy groups with the processing version 020 \citep[briefly described in Appendix~C of][]{Merloni2024}, which is an updated version of 010 processing used for the first data release (DR1). The main updates on the 020 version (internal catalog version 221031) are the improved boresight correction, low-energy detector noise suppression, improved subpixel resolution, and updated pattern and energy tasks. We further reduce the calibrate event files using the using the eROSITA Science Analysis Software System \citep[{\tt eSASS},][]{Brunner2022}\footnote{\url{https://erosita.mpe.mpg.de/dr1/eSASS4DR1/}} with the version id {\tt eSASSusers\_211214} that is the same version used for \citet{Bulbul2024} and \citet{Merloni2024} for DR1. Time variable (solar incident angle dependent) optical light contamination (light leak) is observed in the data from the telescope modules (TMs) 5 and 7, which has a large impact on the calibration of the low-energy band of the spectrum \citep[see][for further details]{Predehl2021,Coutinho2022,Merloni2024}. In this work, we analyze the hot gas properties of galaxy groups whose emission peaks at the soft X-ray band that suffers from the contamination; therefore, we followed a conservative approach and only used the data from TMs 1, 2, 3, 4, and 6, removing the data from TMs 5 and 7 that suffer from contamination due the optical light leak. Furthermore, we obtain a clean event list by applying the standard flag {\tt 0xE000F000} to select all the possible patterns (singles, doubles, triples, and quads) and run the {\tt flaregti} {\tt eSASS} task to have flare filtered good-time-intervals.

\subsection{Imaging analysis}
\label{sec:imaging_analysis}

For the imaging analysis, we use an energy band of 0.3--1.8~keV to maximize the signal-to-noise ratio for a soft X-ray emitting source such as groups (see Appendix~\ref{sec:ebandselection} for the details of this optimization scheme). We extract images, vignetted and non-vignetted exposure maps in this band centered around each group in the catalog using the {\tt evtool} and {\tt expmap} tasks in \texttt{eSASS} with a standard \rosi\ pixel size of 4~arcsec and the {\tt FLAREGTI} option. For the extraction region, we used an image size of $\sim 8r_{500c,\texttt{eSASS}}\times8r_{500c,\texttt{eSASS}}$ that covers well the region from the source center beyond the Virial radius for the local background measurements. The radius, $r_{500c,\texttt{eSASS}}$, was estimated using the flux reported in column {\tt ML\_FLUX\_1} of the eRASS1 X-ray catalog \citep{Merloni2024} and an $L_{\rm X}-M_{500c}$ relation of the \rosi\ Depth Final Equatorial Survey (eFEDS) clusters and groups \citep{Bahar2022, Chiu2022}. These estimates were only used to determine the image size that has a negligible impact on the results. After generating X-ray images and exposure maps, we used the eRASS:4 point source catalog in the 0.2--2.3~keV band to mask or co-fit the point sources in the field of view in the rest of the imaging analysis. Following the same procedure in our eFEDS analysis \citep{Ghirardini2021, Liu2022,Bahar2022}, we masked the faint point sources with {\tt ML\_RATE\_1}~$ < 0.1$~cnts/s) out to the radii where their emission becomes consistent with the background. On the other hand, we co-fit bright point sources ({\tt ML\_RATE\_1}~$> 0.1$~cnts/s) in the surface brightness analysis. In addition to the bright point sources, we also modeled and co-fit the closest extended sources to the central galaxy group to clean the image from contaminating X-ray emission. The remaining ones in the field are conservatively masked out to their $2r_{500c,\texttt{eSASS}}$. Example \rosi images of a bright nearby group (1eRASS~J024933.9-311126) and a group at the median redshift of our group sample (1eRASS~J045547.0-572404) are shown in Fig.~\ref{fig:xray_optical_example}. 

Following a forward modeling approach, we fit the X-ray images using a Bayesian fitting pipeline to deproject the surface brightness emission. We assume a Poisson likelihood for the X-ray counts and sample the likelihood using the {\tt emcee} package  \citep{emcee}  that employs the \citet{GoodmanWeare2010} Affine Invariant Markov chain Monte Carlo (MCMC) technique. The fit is performed to account for the cross-talk between the emission from the nearby co-fitted extended and point sources. For extended sources, we model emissivity using a modified \citet{Vikhlinin2006} profile:

\begin{equation}
\label{eqn:vikhlinin_ne}
\Lambda_{\rm ep}(T, Z)~n_{\rm e}^2(r) = N_{\Lambda, n_{\rm e}^2}~\left( \frac{r}{r_c} \right)^{-\alpha} \left( 1 + \left( \frac{r}{r_c} \right)^2 \right)^{-3\beta+\alpha/2} \left(  1 + \left( \frac{r}{r_s} \right)^3 \right)^{-\epsilon/3},
\end{equation}

\noindent where $\Lambda_{\rm ep}(T, Z)$ is the band-averaged cooling function, $Z$ is the metallicity, and $N_{\Lambda, n_{\rm e}^2}$, $r_c$, $r_s$, $\alpha$, $\beta$, $\epsilon$ are the free parameters of the emissivity profile, normalization, core radius, scale radius, and the power law exponents, respectively. These parameters are allowed to vary in the fits. At each MCMC step, the profile is projected along the line of sight following the equation

\begin{equation}
S_{X} =\frac{1}{4 \pi(1+z)^{4}} \int n_{\rm e} n_{\rm p} \Lambda_{\rm ep}(T, Z) {\rm d}l,
\end{equation}

\noindent where the number density of protons ($n_{\rm p}$) are related to the number density of electrons ($n_{\rm e}$) via  $n_{\rm p} = n_{\rm e}/1.2$ \citep{Bulbul2010}. Then, the projected count rate is convolved with the \rosi\ point spread function (PSF), multiplied with the exposure map, and compared with the masked X-ray image. In addition to the electron density profile parameters, two additional parameters are left free for the central position of the groups, which adds up to eight free parameters for every extended source modeled in the image. For the bright point sources, we only allowed the normalization of their profile to vary while keeping the centroids fixed due to the high positional accuracy of \rosi\ \citep{Brunner2022, Merloni2024}. Lastly, for each image, the background count rates are assumed to be constant across the image, and two more parameters are allowed to be free for the vignetted and unvignetted backgrounds.

\begin{figure*}
\includegraphics[width=0.49\textwidth]{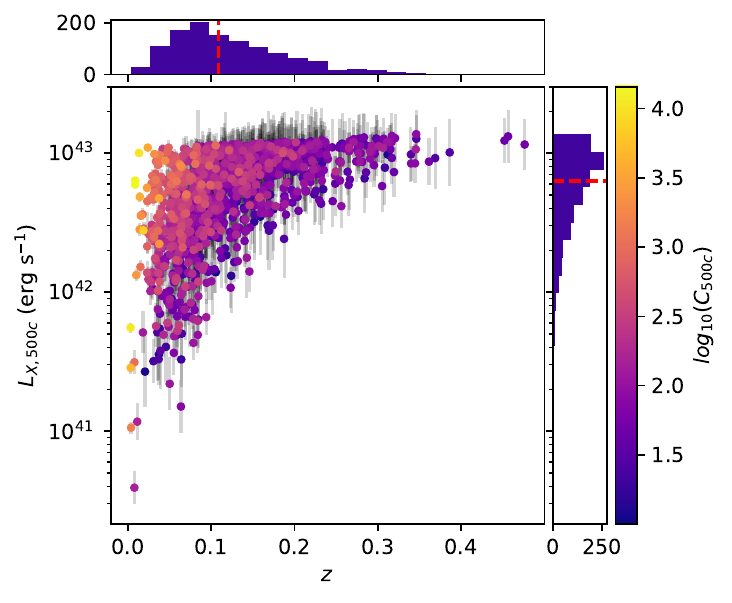} 
\includegraphics[width=0.49\textwidth]{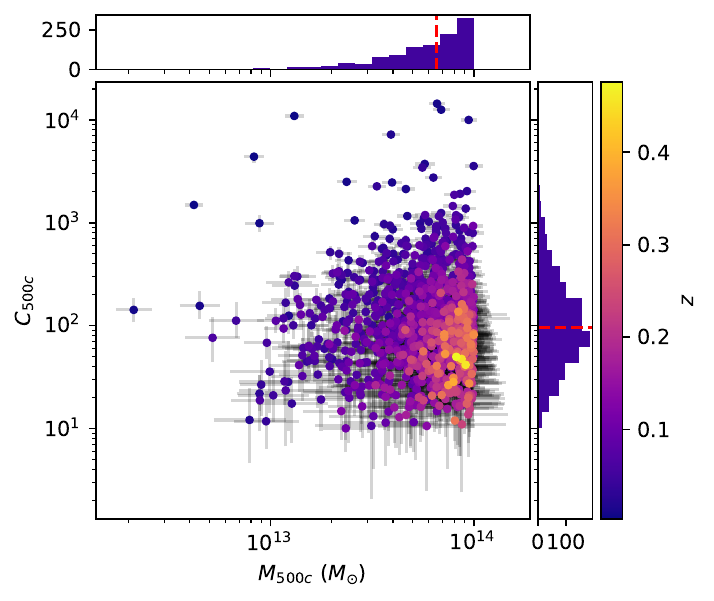} 
\caption{{\it Left:} Soft-band (0.5--2 keV) X-ray luminosity ($L_{X,500c}$) and redshift ($z$) distributions of the galaxy group sample used in this work consisting of 1178 objects. The luminosity of the groups cover a range of $3.9 \times 10^{40} - 1.4 \times 10^{43}$ ergs~s$^{-1}$, and the redshift span of the sample is $0.003-0.48$. {\it Right:} Count ($C_{500c}$) and mass ($M_{500c}$) distributions of the 1178 galaxy group used in this work. Measured counts of galaxy groups range between $10-14380$, and their masses span a range of $2.1 \times 10^{12} -  10^{14}~M_{\odot}$. Median values of the $L_{X,500c}$, $z$, $C_{500c}$ and $M_{500c}$ observables are $6.3 \times 10^{42}$ ergs~s$^{-1}$, 0.11, 95, $6.5 \times 10^{13}~M_{\odot}$ respectively.\label{fig:mass_cn_z_hist}}
\end{figure*}

As an output of this fitting procedure, we obtain the best-fit de-projected emissivity profiles ($\Lambda_{\rm ep}(T, Z)~n_{\rm e}^2(r) $), count-rate profiles, and the associated uncertainties that include the cross-talk between the co-fitted nearby extended and point sources. We show the performance of our pipeline in Fig.~\ref{fig:residual_crprof} for a bright group (1eRASS~J045547.0-572404, the second group in Fig.~\ref{fig:xray_optical_example}) with the images before and after subtracting the emission from the modeled extended and point sources on the left. It is clear from the bottom left figure that after the removal of the modeled profiles, the image is free from any X-ray source, and only noise remains. The surface brightness profile of the observed field (in blue) and the best-fit model (in red) are shown on the right panel of the same figure. The PSF convolved best-fit surface brightness model represents the \rosi\ data well. The peaked emission at large radii (at 500 and ~1000~arcsec) shows the contribution of the modeled point sources to the overall emission, which are successfully modeled and removed from the total source model.

Given the PSF of eROSITA being relatively large, we also ran tests on the robustness of our fitting procedure around the core region ($0.15r_{500c}$) of groups by simulating and fitting synthetic galaxy group observations. The synthetic observations are obtained by first generating galaxy group profiles in a non-parametric way employing a covariance matrix obtained from XXL observations following \citet{2020Comparat}.\footnote{The covariance matrix used to synthesize galaxy group profiles is publically available in \url{https://github.com/domeckert/cluster-brightness-profiles}} These profiles are then convolved with the eROSITA PSF, and the X-ray observations are obtained by creating the Poisson realizations of the PSF convolved surface brightness distributions. Through this procedure, we have fitted 30 simulated groups at a redshift of $z=0.11$ (the median redshift of our sample) and 30 groups at a redshift of $z=0.2$ (85\% of the groups in our sample are at $z<0.2$). As a result of these tests, we found that our fitting procedure is capable of robustly deconvolving the profiles with PSF and recovering the input surface brightness profiles around $0.15r_{500c}$. We also found that because of the PSF smoothing, the recovered profiles of the objects with intrinsically larger surface brightness fluctuations may deviate more from the input simulated profiles; however, at the sample level, these fluctuations cancel out such that our measurements, on average, are unbiased. Furthermore, we have also investigated the possible impact of an undetected central compact source on the surface brightness measurements of the groups in our sample at $0.15r_{500c}$ by comparing the fitted surface brightness profiles of groups with the PSF profile. From this investigation, we find that an undetected point source at the center of a group can only change $n_{\rm e} (0.15r_{500c})$ a few percent, which is within the total error budget of our $n_{\rm e}$ measurements that includes statistical and systematic uncertainties (see Sect.~\ref{sec:systematics} for details on the systematic uncertainties taken into account in this study). Therefore, we conclude that the galaxy groups we use in this work are well extended in the sky such that an undetected point source at the center of the group has little to no impact on the electron density measurements at $0.15r_{500c}$.

\subsection{Estimation of X-ray observables}
\label{sec:ltm_estimation}

The shallow nature of the \rosi\ survey only allows the measurement of the physical properties of a few nearby bright galaxy groups. The \erass\ group sample should be binned into smaller samples to achieve sufficient signal-to-noise for reliably constraining the physical properties of the faint galaxy groups through joint spectral analysis (see Sect.~\ref{sec:grouping_and_the_spectral_analysis}). For an optimal binning scheme, a low-scatter temperature estimator should be used such that groups with similar temperatures can be binned together. Moreover, mass estimates of the galaxy groups are needed to extract spectra within a physically motivated scale radii ($r_{500c}$). For these purposes, we use $L_{\rm X}-M_{500c}$  and $L_{\rm X}-T$ scaling relations and calculate the soft-band ($0.5-2$~keV) X-ray luminosity ($L_{\rm X}$), temperature ($T$), and mass ($M_{500c}$) estimates of the galaxy clusters and groups from the count-rate profiles measured in Sect.~\ref{sec:imaging_analysis}. For a self-consistent treatment of the \rosi\ groups, we employ the $L_{\rm X}-M_{500c}$ and $L_{\rm X}-T$ relations calibrated using the eFEDS observations \citep{Chiu2022, Bahar2022}. 

The selection effects and the mass function need to be accounted for to obtain unbiased estimations of the physical properties of an underlying population from intrinsically scattered scaling relations. For this purpose, we built a Bayesian framework that simultaneously estimates the $L_{\rm X}$ and $T$, $M_{500c}$ observables from the observed count-rate profiles, $\hat{C}_{R}(r)$ (see Sect.~\ref{sec:imaging_analysis} for the details of the count-rate profile measurement procedure). The formulation of the Bayesian estimation framework is as follows. To simultaneously estimate $L_{\rm X}$, $T$ and $M_{500c}$ observables, the joint probability density function, $P(L_{\rm X},T,M_{500c}|D,\theta_{\rm all})$, given the data, $D$, and a set of model parameters, $\theta_{\rm all}$ is needed to be computed. This can be expanded as

\begin{equation}
    P(L_{\rm X},T,M_{500c}|D,\theta_{\rm all}) \equiv P(L_{\rm X},T,M_{500c}| \hat{C}_{R,500c},I,z,\mathcal{H},\theta_{\rm LT},\theta_{\rm LM}),
\label{eqn:joint_prob}
\end{equation}

\noindent where the count-rate within $r_{500c}$ ($\hat{C}_{R,500c}$), detection information of the galaxy group ($I$), redshift ($z$), and sky position ($\mathcal{H}$), represent the data ($D$); and the scaling relations parameters ($\theta_{\rm LT}$ and $\theta_{\rm LM}$) represent the model parameters ($\theta_{\rm all}$). We note that by definition, the $\hat{C}_{R,500c}$ term has an intrinsic dependence on $M_{500c}$ such that $\hat{C}_{R,500c}$ is different for every $M_{500c}$ in the $L_{\rm X}-T-M_{500c}$ parameter space. Taking this into account, our framework allows all the information of the measured count-rate profiles to be included in our analysis rather than count-rate measurements within fixed radii.

Using the Bayes rule, Eq.~\ref{eqn:joint_prob} can be rewritten as
\begin{equation}
\begin{split}
P(L&_{\rm X},T,M_{500c}|\hat{C}_{R,500c},I,z,\mathcal{H},\theta_{\rm LT},\theta_{\rm LM}) = \\
& \frac{P(L_{\rm X},T,M_{500c},\hat{C}_{R,500c},I|z,\mathcal{H},\theta_{\rm LT},\theta_{\rm LM})}{\int \int \int P(L_{\rm X},T,M_{500c},\hat{C}_{R,500c},I|z,\mathcal{H},\theta_{\rm LT},\theta_{\rm LM}) dL_{\rm X} dT dM_{500c}}.
\end{split}
\label{eqn:joint_prob_bayes}
\end{equation}

Furthermore, one can expand the common term in the numerator and the denominator as 

\begin{equation}
\begin{split}
P(L_{\rm X},T,M_{500c},\hat{C}_{R,500c},I|z,\mathcal{H},\theta_{\rm LT},&\theta_{\rm LM}) \approx   \\ 
 P(\hat{C}_{R,500c}|L_{\rm X},T,M_{500c}&,z)  P(I|L_{\rm X}, z, \mathcal{H})  \\ \times P(L_{\rm X},&T|M_{500c}, z,\theta_{\rm LT},\theta_{\rm LM}) P(M_{500c}|z),  
\end{split}
\label{eqn:joint_prob_bayes_term_expanded}
\end{equation}

\noindent where the first term, $P(\hat{C}_{R,500c}|L_{\rm X}, T, M_{500c},z)$, stands for the measurement uncertainty of the count-rate. The conditional probability distribution for this term is obtained by first calculating the true count rate ($C_{R,500c}$) for every point in the $L_{\rm X}-T$ parameter space by assuming the source emitting an unabsorbed {\sc APEC} \citep{Smith2001} spectrum in {\tt Xspec} \citep{Arnaud1996} at a redshift $z$ with an abundance of $0.3Z_{\odot}$, a temperature of $T$ and a luminosity of $L_{\rm X}$. Then the true count-rates ($C_{R,500c}$) are compared with the observed count-rates calculated ($\hat{C}_{R,500c}$) at every $r_{500c}$ value in the mass parameter space ($M_{500c}$) and the value of the conditional probability is obtained. The $P(I|L_{\rm X}, z, \mathcal{H})$ term in Eq.~\ref{eqn:joint_prob_bayes_term_expanded} is the selection function term that is a function of soft-band X-ray luminosity, redshift, and sky position where the sky position includes the local background surface brightness, exposure, and the neutral hydrogen column density information. 

The selection function is obtained by simulating the \rosi\ X-ray All-Sky observations using the baryon painting method \citep{2019Comparat, 2020Comparat} and applying the same routines of the {\tt eSASS} source detection pipeline to construct one-to-one correspondence of the catalogs and selection \citep{Seppi2022, Clerc2024}. The $P(M_{500c}|z)$ term is the mass function term for which the analytical formulation of \citet{tinker08} is used in this work. Lastly, the $P(L_{\rm X}, T|M_{500c}, z,\theta_{\rm LT},\theta_{\rm LM})$ term is the intrinsically scattered scaling relation term that gives the $L_{\rm X}$ and $T$ distributions at a given mass and redshift.

Ideally, one would use a jointly fit, intrinsically scattered  $L_{\rm X}-T-M_{500c}$ scaling relation for the $P(L_{\rm X},T|M_{500c}, z,\theta_{\rm LT},\theta_{\rm LM})$; however, there is no such relation in the literature yet that is calibrated by taking into account the selection effects and covers a similar mass range with eROSITA. For this reason, we expanded this term as

\begin{equation}
    P(L_{\rm X},T|M_{500c}, z,\theta_{\rm LT},\theta_{\rm LM}) \approx P(T | L_{\rm X},z,\theta_{\rm LT}) P(L_{\rm X} | M_{500c},z,\theta_{\rm LM})
\end{equation}

\noindent 
and used the \citet{Bahar2022} $L_{\rm X}-T$ and \citet{Chiu2022} $L_{\rm X}-M_{500c}$ relations that are calibrated for \rosi\ by taking into account the selection effects. The $P(T | L_{\rm X},z,\theta_{\rm LT})$ term was obtained from $P(L_{\rm X}|T,z,\theta_{\rm LT})$ using Bayes theorem:

\begin{equation}
    P(T | L_{\rm X},z,\theta_{\rm LT}) = \frac{P(L_{\rm X}|T,z,\theta_{\rm LT})P(T|z)}{\int P(L_{\rm X}|T,z,\theta_{\rm LT})P(T|z) dL_{\rm X}},
\end{equation}
where self-consistently, the same temperature function is used for the $P(T|z)$ term as in Eq.~6 in \citet{Bahar2022}.

\begin{figure*}
\includegraphics[width=0.99\textwidth]{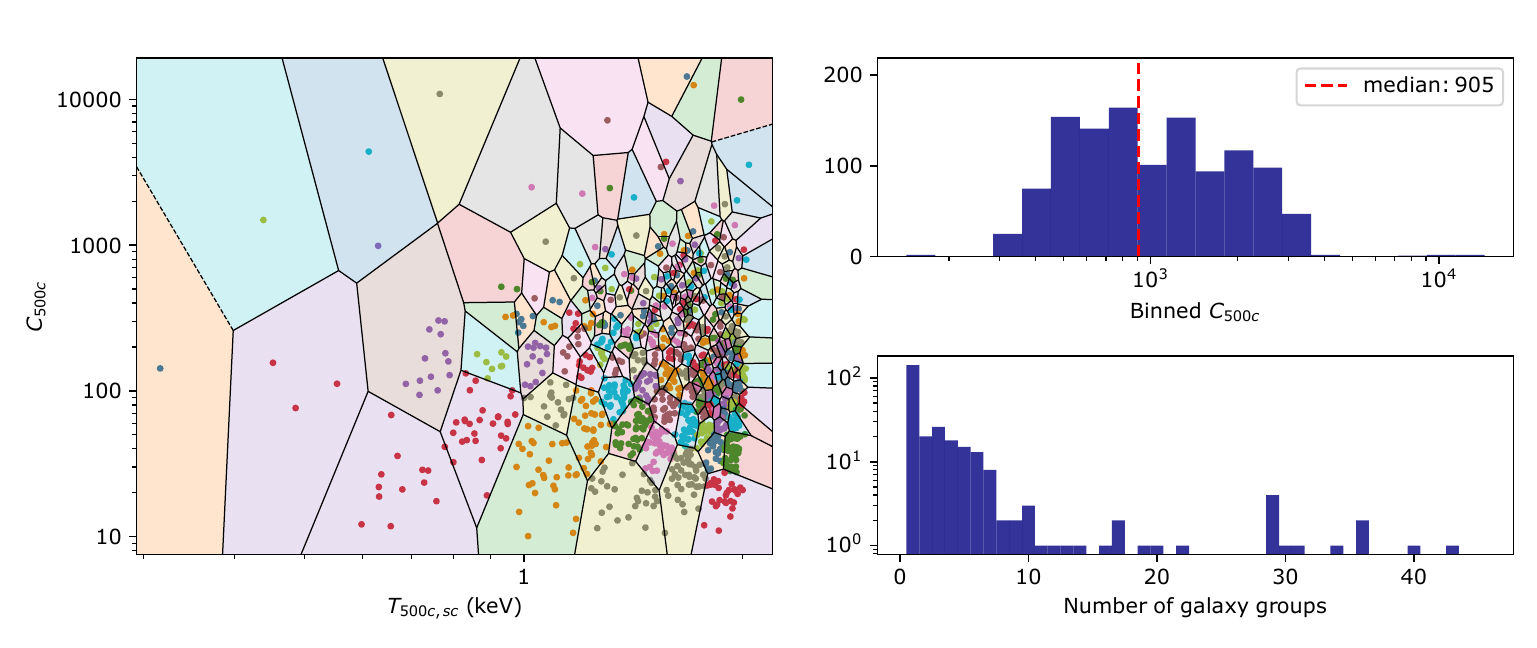} 
\caption{{\it Left:} Voronoi binning scheme used for grouping the sample obtained from the distribution of count (within $r_{500c}$) measurements ($C_{500c}$) and scaling relation based temperature estimates ($T_{500c,sc}$). {\it Top right:} Histogram of total counts in 271 Voronoi bins with a median of 905 counts. {\it Bottom right:} Histogram of galaxy groups in Voronoi bins.\label{fig:voronoi_grouping}}
\end{figure*}

As a final step, we substitute the terms in Eq.~\ref{eqn:joint_prob_bayes_term_expanded} to Eq.~\ref{eqn:joint_prob_bayes} and calculate the joint probability density function, $P(L_{\rm X},T,M_{500c}|\hat{C}_{R,500c},I,z,\mathcal{H},\theta_{\rm LT},\theta_{\rm LM})$, for each galaxy group. Subsequently, we marginalize over the nuisance parameters and obtain $L_{\rm X}$, $T$, and $M_{500c}$ estimates given the data and the scaling relations parameters.\footnote{For example, the marginal probability distribution of temperature is calculated as $P(T | \dots) = \iint P(L_{\rm X}, T, M_{500c} | \dots) dL_{\rm X} dM_{500c}$ and the temperature corresponding to the 50th percentile of the marginal probability distribution is used as the point estimate.} We provide the distributions of mass, temperature, soft-band ($0.5-2$~keV) X-ray luminosity estimates obtained through this Bayesian framework along with the distributions of redshift and count ($0.3-1.8$~keV) of the final galaxy group sample in Figs.~\ref{fig:mass_z_hist}, \ref{fig:mass_cn_z_hist} and \ref{fig:voronoi_grouping}.

The estimated $M_{500c}$ are then converted to $r_{500c}$ and $r_{2500c}$ by assuming an average dark matter concentration of $c_{500c}=r_{500c}/r_{s,\rm NFW}=4.2$\footnote{$r_{s,\rm NFW}$ is the characteristic radius of the Navarro–Frenk–White (NFW) profile and a concentration ($c_{500c}$) of 4.2 corresponds to a $r_{2500c} /r_{500c}$ ratio of 0.465.} and scaling the $r_{500c}$ estimates accordingly (S09). These characteristic radii are employed to determine the spectral extraction region ($r<r_{500c}$) and serve as characteristic radii ($0.15r_{500c}$, $r_{2500c}$, $r_{500c}$) for the entropy measurements. 

The net effect of accounting for selection effects and the mass function when estimating $L_{\rm X}$, $T$, and $M_{500c}$ from scaling relations depends on two factors: the scatter in the scaling relations and the interplay between the selection and mass functions. With zero scatter, there's a one-to-one relationship between observables, allowing straightforward conversions. As the scatter of the relation increases, the scaling relation estimates that ignore the selection effects will be more vulnerable to being biased. Furthermore, the net effect also depends on the shapes of the selection and mass functions. The interplay between the selection function and the mass function across the $L_{\rm X}-T-M_{500c}$ parameter space is often not trivial; however, to the first order, if we consider X-ray selection as a redshift dependent $L_{\rm X}$ cut, not accounting for selection and mass functions would lead to both $T$ and $M_{500c}$ being overestimated.

%
\subsection{Grouping and the spectral analysis}
\label{sec:grouping_and_the_spectral_analysis}

Measuring temperature through X-ray spectroscopy requires considerably more photons than measuring surface brightness properties with imaging analysis. Given the shallow nature of the \rosi\ All-Sky Survey, the photon counts of most of the galaxy groups in our sample are insufficient for temperature measurements, even though the flux or luminosity of these objects can be reliably measured from X-ray images. For instance, more than half of the groups in the sample, shown on the right panel of Fig.~\ref{fig:mass_cn_z_hist}, have fewer than 100 counts within the $0.3-1.8$~keV band, which is not sufficient for measuring their temperature reliably.

The two canonical ways to overcome the problem of insufficient photon counts for spectral analysis are co-fitting or stacking. A plethora of examples of both techniques exist in the literature. For example, \citet{McDonald2014} co-fit radially extracted spectra of 80 South Pole Telescope (SPT) selected massive clusters, while \citet{Bulbul2014} and \citet{Zhang2024} stacked megaseconds of \xmm and \rosi\ spectra respectively to achieve a high signal-to-noise level and reveal faint spectral features. In this work, we employ the co-fitting technique to maintain the spectral information of individual groups that would be averaged out when stacked. This method is the most suitable for the primary goal of this work. Compared to stacking, this approach is computationally expensive; however, with the recently developed high-performance Central Processing Units (CPUs) and the improvements in parallel computing, we are able to employ the co-fitting technique in this work. 

We grouped the sample such that the IGrM temperatures of the galaxy groups are similar in each bin. Moreover, we required the statistical constraining power (photon counts) of the groups in the same bin to be similar to each other to avoid the source with the highest count from biasing the measurements. In other words, our aim was to minimize the temperature and photon count variation -- $\Delta T_{500c} \sim std(T_{500c})$\footnote{The notation $std(X)$ represents standard deviation of X.} and $\Delta C_{500c} \sim std(C_{500c})$ -- in each bin while trying to achieve a sufficient signal-to-noise ratio. To achieve this, we grouped the sample using the Voronoi binning technique \citep{Cappellari2003}. To apply the tessellation technique, we pixelated our temperature proxy $T_{500c,sc}$ (surface brightness inferred temperature estimate; see Sect.~\ref{sec:ltm_estimation} for the details) and count measurements $C_{500c}$ such that each pixel was occupied by only one galaxy group. 
The axes are then re-scaled, and the resulting image is given to the {\tt Vorbin} package, the Python implementation of the Voronoi binning technique. The free parameters, namely, the axes scaling factors and the target S/N, are fine-tuned until the temperature variation ($\Delta T_{500c}$) and the photon count variation ($\Delta C_{500c}$) in the Voronoi bins are sufficiently small. The final binning scheme, shown in Fig.~\ref{fig:voronoi_grouping}, is achieved using an S/N target, $S/N = 22.36$ (equivalent to 500 counts). We further present the distributions of the total counts and the number of galaxy groups in the Voronoi bins in Fig.~\ref{fig:voronoi_grouping}. Using this binning scheme, we obtained 271 bins with a median count of 905, sufficient for obtaining reliable spectroscopic temperature measurements at $T<2$ keV for each bin. We note that the binning scheme can be slightly different if a different target S/N or axis scaling factors are chosen; however, the impact of the chosen binning scheme is negligible on the final results as long as the resulting $\Delta T_{500c}$ and $\Delta C_{500c}$ are similar.

\begin{figure}
\includegraphics[width=0.49\textwidth]{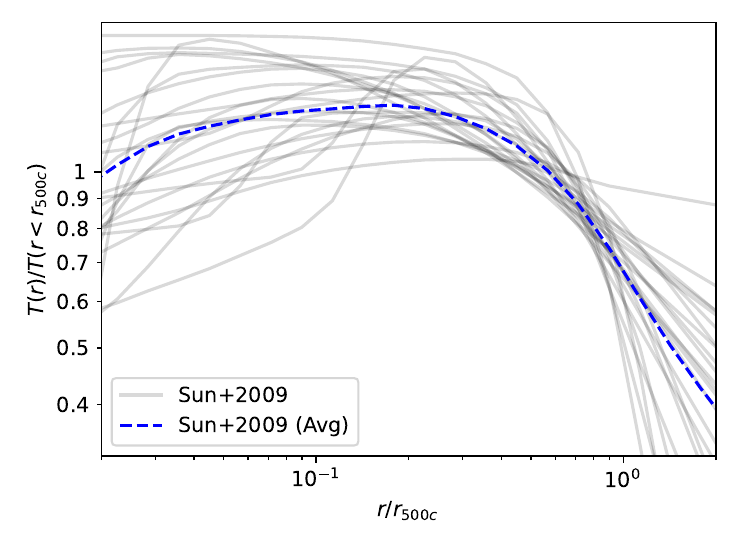}
\caption{Normalized temperature profiles, $T(r)/T(r<r_{500c})$, of 23 groups (gray) presented in S09 and their average (blue). The blue dashed line provides an average conversion ratio between the temperature profile, $T(r)$, and the characteristic temperature measurements, $T(r<r_{500c})$. \label{fig:deproj_temp_sun}}
\end{figure}

After the grouping, the source and local background spectra of the galaxy groups in our sample are extracted using the {\tt eSASS} task, {\tt srctool}. The source spectra are extracted from the circular regions centered around the galaxy group and have a radius of $r_{500c}$ (see Sect.~\ref{sec:ltm_estimation} for the details of the $M_{500c}$/$r_{500c}$ estimation procedure). Similarly, the local background spectra are extracted from annuli that are centered around the galaxy group and have a radial range of $4r_{500c} < r < 6r_{500c}$. The best-fit count-rate profiles are used to determine the masking radius of the bright point sources and nearby extended objects co-fitted during the imaging analysis. The remaining point sources and nearby extended sources within the extraction region are masked as described in Sect.~\ref{sec:imaging_analysis}.

We extract ancillary response files (ARFs) and redistribution matrix files (RMFs) using the {\tt srctool} task for the background and the source region in different settings to be assigned to various components of the source and background models. The ARF assigned to the source component is extracted with the {\tt exttype=BETA} and {\tt psftype=2D\_PSF} settings to consider the energy-dependent PSF and vignetting corrections. Over the extraction regions, the flux distribution of the vignetted X-ray background is assumed to be flat, and the {\tt exttype=TOPHAT} and {\tt psftype=NONE} settings are used for extracting the ARFs assigned to the vignetted X-ray background components of the source and background regions. 

Similar to our eFEDS analysis \citep{Ghirardini2021, Liu2022, Bahar2022, Bulbul2022}, the local background model consists of two major components: particle-induced instrumental background \citep[see][for further details]{Bulbul2020, Freyberg2020} and X-ray background including the Galactic foreground, and unresolved point sources in the sky. The total model includes a spectral model component with an absorbed thermal component. The spectral analysis was performed using {\tt PyXspec}, the Python interface of the standard X-ray spectral analysis package {\tt Xspec} \citep[version 12.12.1,][]{Arnaud1996}, which employs the {\tt AtomDB} atomic database \citep[version 3.0.9,][]{2012Foster}. The {\tt Xspec} model of the  X-ray foreground consists of an unabsorbed {\sc APEC} \citep{Smith2001} for the local hot bubble \citep[$T \sim 0.084$~keV]{Yeung2023}, two absorbed {\sc APEC}s for the hot and cold components of the galactic halo \citep[$T \sim 0.49$ and $0.157$~keV respectively]{Ponti2023, Bulbul2012}. To model the cosmic X-ray background, we use an absorbed power-law for the unresolved AGN \citep[$\Gamma = 1.45$]{Cappelluti2017}. For the shape of the instrumental background, we use the best-fitting model of \citet{Yeung2023} obtained by calibrating the filter wheel closed (FWC) data \citep[see Appendix A.1. and A.2. of][for the details of the modeling of FWC data]{Yeung2023}. This instrumental background component is folded with unvignetted ARF \citep{Freyberg2020}, while the cosmic X-ray background and Galactic foreground are folded with the respective vignetted ARF in the fits.

\begin{figure}
\includegraphics[width=0.49\textwidth]{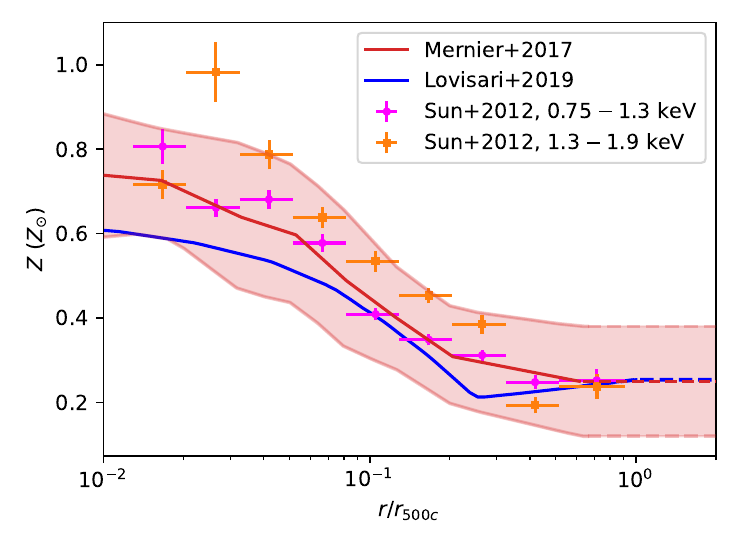} 
\caption{Average metalicity profiles of galaxy groups reported by \citet{Mernier2017} and \citet{Lovisari2015} and the stacked metallicity profiles of \citet{Sun2012} for galaxy groups in two temperature bins ($T = 0.75-1.3$~keV and $0.75-1.3$~keV).\label{fig:metallicity_prof}}
\end{figure}

To account for the X-ray absorption, we use the {\sc TBABS} \citep{Wilms2000} interstellar medium (ISM) absorption model in {\tt Xspec} \citep{Arnaud1996}. We use the HI4PI survey \citep{HI4PI2016} for calculations of the hydrogen column density ($n_{\rm H}$). The $n_{\rm H}$ values at the positions of eRASS1 galaxy groups are relatively low because of their locations at higher Galactic latitudes; therefore, using the total hydrogen column density ($n_{\rm H,tot}$) rather than the neutral hydrogen column density ($n_{\rm H, I}$) has a negligible impact on our results at the sample level\footnote{To confirm this statement, we ran tests to quantify the impact of using $n_{\rm H,tot}$ \citep{Willingale2013} instead of $n_{\rm H,I}$ \citep{HI4PI2016} on the sample averaged quantities presented in Sect.~\ref{sec:results}, $S(r)$ and $T(r<r_{500c})$. The tests showed that the choice has overall a negligible impact ($< 3\%$) compared to the systematic error budget of the average quantities (see Sect.~\ref{sec:systematics}).} as also noted in \citet{Bulbul2024}. We use the solar abundances of \citet{Asplund2009} when measuring the metallicity of the groups. We use C-statistic \citep{Cash1979} for the statistical interpretation of our spectra that provides unbiased estimates of the model parameters at the low and high count regimes \citep{Kaastra2017}. We employ the co-fitting technique for the spectral analysis. This required us to explore likelihoods with relatively high dimensional parameter space. For this purpose, we chose to employ the MCMC fitting technique. {\tt Xspec} has a built-in MCMC sampler; however, the amount of control it allows the user over the priors is limited. For this reason, we employ the widely used MCMC sampler {\tt emcee} \citep{emcee} rather than the built-in {\tt Xspec} sampler to explore high dimensional likelihoods. We achieved this by developing an interface that allows cross-talk between {\tt PyXspec} and {\tt emcee} and updates the model parameters at every MCMC step accordingly.

Galaxy groups are low-mass objects with relatively low plasma temperatures ($T <2$~keV) due to their shallower potential wells. They share this low-temperature parameter space with other background/foreground components, such as the cold ($T\sim0.157$~keV) and hot ($T\sim0.49$~keV) components of the galactic halo \citep{Ponti2023}. This results in degeneracies between the source and background/foreground components at the low count regime. At a given energy band, it is relatively easy to separate the source and background components using the 2D distribution of photons through imaging analysis since we expect the local background rate to be relatively flat, whereas the source emission roughly follows a projected Vikhlinin profile \citep{Vikhlinin2006}. In this work, we make use of this fact and combine the spatial and spectral information of photons following a novel approach with the aim of lifting the aforementioned degeneracies the best we can. We achieve this by using the observed count-rates ($0.3 - 1.8$~keV) measured through imaging analysis as priors in the spectral analysis in our pipeline that combines {\tt PyXspec} and {\tt emcee}.

To obtain the average temperatures of the binned galaxy groups (see the first paragraph of this section for the details of the grouping), we only link the temperature parameter of the ({\sc APEC}) model and co-fit all the source and background spectra of the galaxy groups in each Voronoi bin. In total, $2 \times N_{i,\rm gr}$ spectra are co-fit (one spectrum for the source region and one for the background region) where $N_{i,\rm gr}$ is the number of galaxy groups in the $i$'th Voronoi bin. During the fitting, the temperatures, the normalizations of the X-ray background components, and the normalizations of the unvignetted particle background components are allowed to be free with flat priors in the logarithmic parameter space. 

Many studies in the literature show strong degeneracy between the temperature and metallicity measurements for galaxy clusters and groups hosting multi-phase gas. This manifests as the so-called 'Fe bias' \citep{Buote1998,Buote2000,Gastaldello2021,Mernier2022} where temperature and abundance measurements change depending on the number of gas components fitted. In our work, we take this effect into account by allowing the metallicities of each galaxy group to be free with a Gaussian prior centered around 0.3~$Z_{\odot}$ with a standard deviation of 0.025 while measuring average temperatures. The normalization of the source emission of each galaxy group and the co-fit temperature are left free with log-uniform priors. Consequently, following the spectral co-fitting analysis procedure described above, we obtained 271 average temperature measurements within $r_{500c}$ for each Voronoi bin shown in Fig.~\ref{fig:voronoi_grouping}.

\subsection{Electron density, temperature, and entropy profiles}

From the imaging analysis described in Sec.\ref{sec:imaging_analysis}, we obtain deprojected emissivity profiles of all the 1178 galaxy groups. The spectral analysis, described in Sec.\ref{sec:grouping_and_the_spectral_analysis}, yields average temperature measurements of the 271 galaxy group bin within $r_{500c}$. The electron number density profile measurements of the IGrM have a non-negligible dependence on the assumptions on the temperature and metallicity profiles. Furthermore, temperature profiles of the galaxy groups are needed for obtaining entropy profiles. For most binned groups, only a single average temperature and metallicity measurement can be achieved within $r_{500c}$ due to low S/N \rosi\ data. We overcome this limitation and incorporate the impact of temperature variation on the thermodynamic properties as a function of radius, as described below.

To account for the radial temperature variation, we first determine the average shape of the 3D temperature profile of groups ($T(r)/T(r<r_{500c})$) using the temperature profile measurements of the tier 1 and 2 groups presented in S09 (see Sect.~\ref{sec:sys_kT_metal} for the details). The average and individual shapes of the $T(r)/T(r<r_{500c})$ profiles are shown in Fig.~\ref{fig:deproj_temp_sun}. We then rescale the average shape with the integrated temperature measurements and obtain the average temperature profiles for each binned group. The temperature profile measurements presented in S09 for a sample of 43 groups are obtained by analyzing deep \chandra observations; therefore, the overall shape of the profiles is relatively well-constrained. Following this procedure, we obtain the average temperature profiles of the 271 galaxy group bins. We note that our approach of obtaining temperature profiles of groups is equivalent to fixing the shape of an assumed temperature profile and fitting spectra by allowing the normalization of the profile to be free. We further note that the observed temperature measurement discrepancy between telescopes \citep[e.g.,][]{Liu2023} does not affect our work given that the temperature measurements of \rosi for galaxy groups agree very well with the \chandra and \xmm temperatures \citep{Migkas2024}.

For the metallicity profile, we consider the following studies in the literature that have reasonably large galaxy group samples: \citet{Sun2012}, \citet{Mernier2017}, and \citet{Lovisari2019}. We find that their measurements agree relatively well within the scatter of the metallicity profile reported in \citet{Mernier2017} (see Fig.~\ref{fig:metallicity_prof}). For this reason, we adopt the median \citet{Mernier2017} metallicity profile ($Z_{M17}$) for our measurements and consider the scatter of their profile as our systematic uncertainty. We quantify the impact of this uncertainty on the thermodynamic profile measurements ($n_{\rm e}(r)$ and $S(r)$) and consider the resulting difference as part of the overall error budget (see Sect.~\ref{sec:sys_kT_metal} for the details of the assumed metallicity profile and quantification of the impact of this choice). 

We then calculate the electron density profiles from our deprojected emissivity measurements by constructing {\tt APEC} models in {\tt Xspec} \citep[similarly done in][]{Liu2022} having the temperatures equal to the temperature profiles of the binned groups and the metallicities equal to the assumed metallicity profile at all radii out to $2r_{500c}$. During the $n_{\rm e}$ profile calculation, in addition to the errors of the imaging analysis, the uncertainties of the average metallicity and temperature measurements are propagated as well using the MCMC chains of the spectral analysis, taking into account the covariances. Then, the electron density profiles of the objects within each Voronoi bin are averaged to get the average electron density profiles of the binned groups. 

Lastly, the entropy profiles of the binned groups are obtained by combining the average electron density and the temperature profiles using the equation below:

\begin{equation}
S(r)= T(r)/n_{\rm e}(r)^{2/3}.
\end{equation}
The entropy profiles are then sampled at three characteristic radii, and the final entropy measurements of 271 galaxy group bins are obtained. The full shape of the entropy profiles of binned groups, along with the other thermodynamic profiles such as electron density and pressure ($P = n_{\rm e} T$), will be presented in Bahar et al. (in prep.).

Besides the systematics resulting from the metallicity profile, we also consider other major systematics that have a non-negligible impact on the thermodynamic properties measured in this work. We discuss and quantify the impact of these systematics on our measurements in Sect.~\ref{sec:systematics} and take them into account as part of the total error budget when we draw conclusions in the next section.

\begin{figure}
\includegraphics[width=0.49\textwidth]{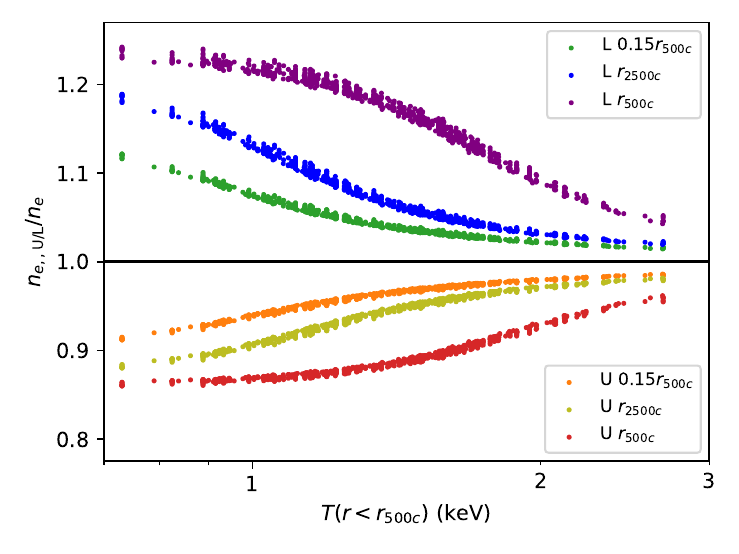} 
\caption{Ratio of the electron densities obtained by assuming low/up-scattered \citet{Mernier2017} metallicity profiles to the electron densities obtained by the median \citet{Mernier2017} metallicity profile at three radii, $0.15r_{500c}$, $r_{2500c}$, and $r_{500c}$ as a function of characteristic temperature, $T(r<r_{500c})$. Green, blue, and purple data points represent the ratio between the electron densities obtained by assuming the lower envelope ($n_{{\rm e},L}$) of the red shared area and the dark red median line ($n_{\rm e}$) in Fig.~\ref{fig:metallicity_prof}. Orange, yellow, and red data points represent the ratio between the electron densities obtained by assuming the upper envelope of the red shared area ($n_{{\rm e},U}$) and the dark red median line ($n_{\rm e}$) in Fig.~\ref{fig:metallicity_prof}.\label{fig:ne_ratio_t}}
\end{figure}
\section{Assumptions, corrections, and systematics}
\label{sec:systematics}

Having fair comparisons between the thermodynamic properties of groups observed with different X-ray observatories and simulations is a challenging task as various systematics should be taken into account in the measurements, such as the systematic uncertainties on the metallicity and temperature profiles, systematic uncertainties on the group masses, the flux calibration mismatches between instruments and the systematics resulting from the use of different atomic database versions. In this section, we provide a list of assumptions, corrections, and systematics taken into account in this work, along with our approach to account for them. We also list a summary of the description and implementation of the assumptions, corrections, and systematics in Table~\ref{tab:table_of_systematics}.

\begin{figure}
\includegraphics[width=0.49\textwidth]{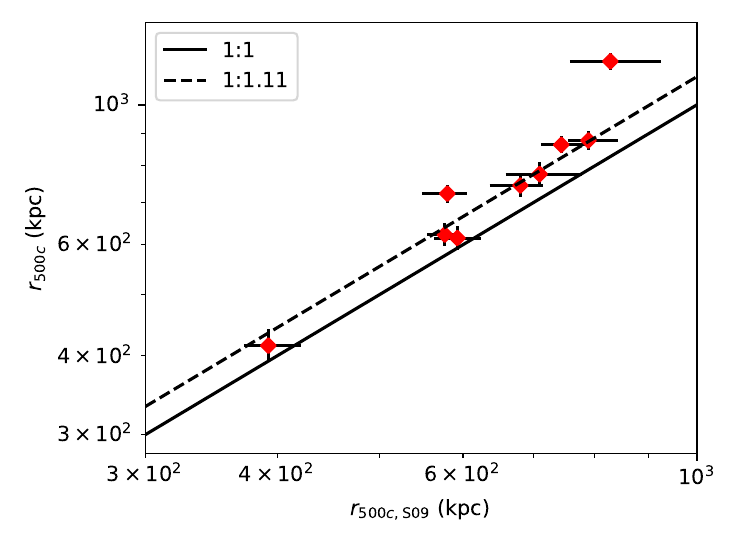}
\caption{Comparison between the scaling relation based $r_{500c}$ estimates of the crossmatched galaxy groups obtained in this work (y-axis: $r_{500c}$) with the estimates of S09 (x-axis: $r_{500c, \rm S09}$) obtained by assuming hydrostatic equilibrium.\label{fig:r500_comp}}
\end{figure}
\begin{figure*}
\includegraphics[width=0.3\textwidth]{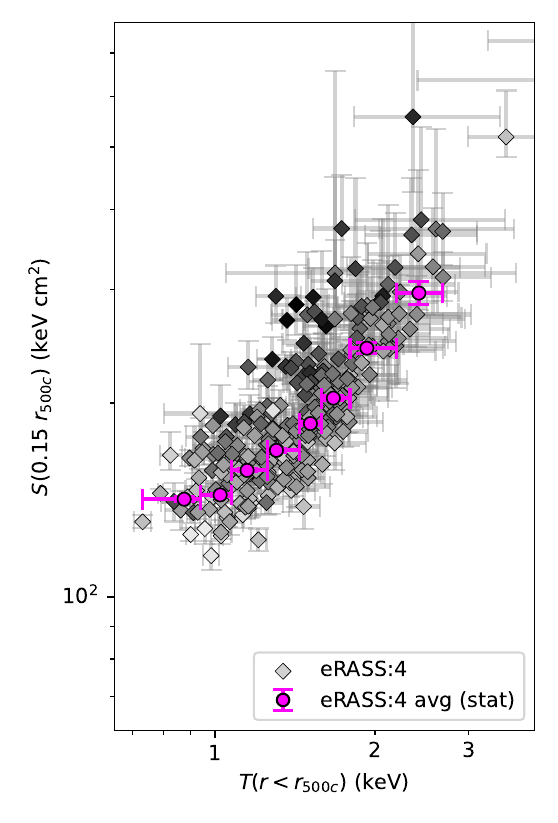} 
\includegraphics[width=0.303\textwidth]{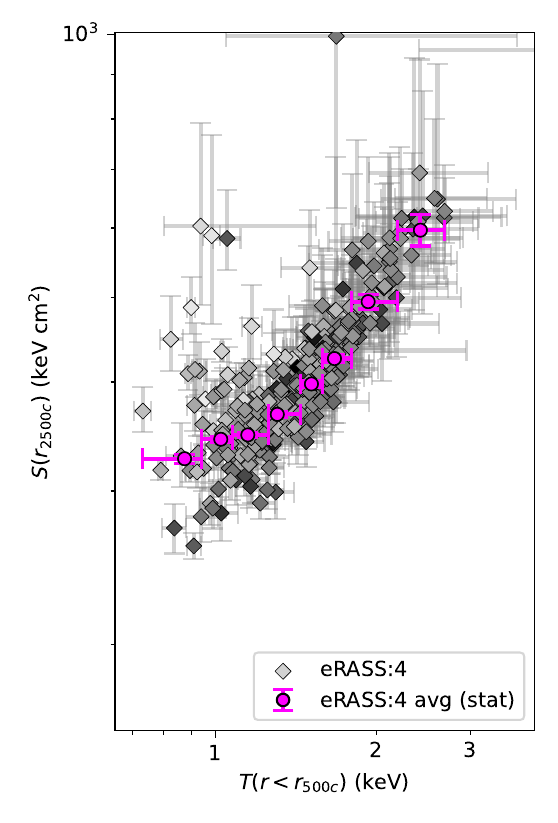} 
\includegraphics[width=0.3823\textwidth]{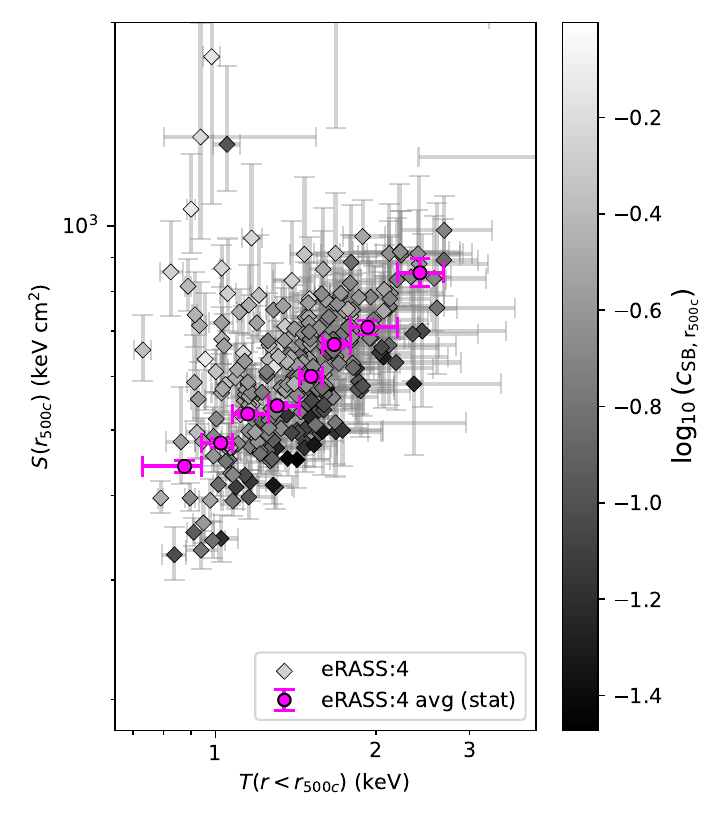} 
\caption{Average entropy measurements of binned groups (diamonds) at three radii ($0.15r_{500c}$, $r_{2500c}$, $r_{500c}$) as a function of characteristic temperature, $T(r<r_{500c})$. The colors of the diamond data points represent the average concentration ($c_{{\rm SB},r_{500c}} = SB(r<0.1r_{500c})/SB(r<r_{500c})$) of the groups within the corresponding Voronoi bin. Error-weighted averages of the diamond data points are shown in magenta.\label{fig:Sall_T_cc}}
\end{figure*}
\subsection{Assumptions on temperature and metallicity profiles}
\label{sec:sys_kT_metal}

Accounting for the temperature and metallicity radial variation is key to having reliable thermodynamic profiles. Shallow survey observations and low signal-to-noise data of most groups in our sample are insufficient to measure temperature profiles reliably. There are various studies in the literature on the average temperature profile of the hot gas in clusters \citep[e.g.,][]{McDonald2014,Ghirardini2019}, while the studies focusing on the shape of the average temperature profile of groups with a large enough sample are limited. An in-depth study of 43 nearby galaxy groups with deep \chandra observations by S09 (among which 23 of them have good temperature constraints out to $r_{500c}$) is one of the few studies we compared within this work. In this work, we used the temperature profile measurements of these 23 groups to get the average shape of the temperature profile of groups. To get the average shape, we first calculated the characteristic temperatures of the groups, $T(r<r_{500c})$, by projecting and integrating all the temperature profiles within a cylindrical volume of radius $r_{500c}$. We achieved this by following the \citet{Mazzotta2004} weighting and projection formulas
\begin{equation}
w = n_{\rm e} n_{\rm p} T^\alpha
\end{equation}
and
\begin{equation}
T_{\rm sl} = \frac{\int w T dV}{\int w dV},
\end{equation}
where for $\alpha$ we used $0.76$, which is calibrated for \rosi\ \citep{ZuHone2023}. Furthermore, we used our average group electron density profile for $n_{\rm e}$ (Bahar et al., in prep.). We then divided the temperature profiles with the characteristic temperatures and obtained the normalized temperature profiles, $T(r)/T(r<r_{500c})$. Lastly, we took the average of these profiles and renormalized them to get the average shape of the temperature profiles of groups. The average and individual normalized profiles of 23 groups are shown in Fig.~\ref{fig:deproj_temp_sun}.

Unlike clusters, the band-averaged cooling function of groups has a strong metallicity dependence because of the significant contribution from the line emission at temperatures, $T<2$~keV. Therefore, the radial change in metallicity from the center to the outskirts should be accounted for to calculate electron density profiles accurately.

During the last decade, the shape and the strength of emission lines have significantly changed in most commonly used plasma emission codes (see Sect.~\ref{sec:sys_atomic_database}) that had a strong influence on the metallicity measurements of groups \citep[e.g.,][]{Mernier2018}. For this reason, it is important to use the most recent publications and account for uncertainties in metallicity profiles in the systematics error budget. Among the metallicity profile measurements in the literature, the results presented in \citet{Sun2012}, \citet{Mernier2017} and \citet{Lovisari2019} stand out as the most recent studies with moderately large galaxy group samples with sufficiently deep observations. Fig.~\ref{fig:metallicity_prof} presents the stacked metallicity profiles of \citet{Sun2012} and the average metallicity profiles of \citet{Mernier2017} and \citet{Lovisari2019} that are renormalized based on the iron abundance ratio of \citet{Asplund2009}. In \citet{Sun2012}, the author reports stacked abundance profiles of 39 galaxy groups in three temperature bins ($0.75-1.3$, $1.3-1.9$ and $1.9-2.7$~keV). In this work, we consider only the results of the first two temperature bins ($0.75-1.3$~keV and $1.3-1.9$~keV), which are relevant to our sample that has a median temperature of $T(r<r_{500c})=1.45$~keV.

\begin{table*}
\centering
\caption{Positions, redshifts, and $r_{500c}$ estimates of the crossmatched groups with S09.}
\label{tab:r_500}
\begin{tabular}{lcccccc}
\toprule\midrule
eROSITA ID   & Literature name & RA$^\ast$ & DEC$^\ast$ & $z^\ast$ & $r_{500c}$$^{\dagger}$ \\
(1eRASS)  &  & (deg) & (deg) & & (kpc) \\
\midrule
J120427.3+015346 & MKW4 & 181.114 & 1.896 & 0.0203 & 723 \\
J093325.7+340302 & UGC 5088 & 143.357 & 34.051 & 0.0269 & 415 \\
J120638.9+281024 & NGC 4104 & 181.662 & 28.173 & 0.0283 & 623 \\
J110943.5+214545 & A1177 & 167.432 & 21.763 & 0.0322 & 615 \\
J054006.9-405004 & ESO 306-017 & 85.029 & -40.835 & 0.0368 & 864 \\
J000313.1-355607 & A2717 & 0.805 & -35.935 & 0.0500 & 878 \\
J102212.8+383136 & RXJ 1022+3830 & 155.553 & 38.527 & 0.0544 & 745 \\
J231358.6-424338 & AS1101 & 348.494 & -42.727 & 0.0557 & 1172 \\
J131214.0-005825 & A1692 & 198.059 & -0.974 & 0.0843 & 776 \\
\bottomrule
\end{tabular}
\tablefoot{
\\
$^\ast$ RA, DEC, and $z$ of the listed groups are taken from \citet{Bulbul2024}. \\
$^\dagger$ Scaling relation based $r_{500c}$ estimates used in this work are obtained by fully accounting for the selection and the mass functions. \\
}
\end{table*}

Overall, the average metallicity profile reported in \citet{Mernier2017} lies between the \citet{Sun2012} and \citet{Lovisari2019} measurements and the \citet{Sun2012} measurements in the $1.3-1.9$~keV temperature bin lies above the \citet{Mernier2017} profile and the average measurements of \citet{Lovisari2019} lie below the \citet{Mernier2017} profile. When calculating the thermodynamic properties, the differences in metallicity measurements must be accounted for as systematics because of the strong dependence of emissivity on metallicity at group scales. Given the large spread of metallicity measurements, we take the average profile of \citet{Mernier2017} as our default profile and conservatively consider the shaded area as the systematics of the average profile measurements. To account for the impact of the choice of average metallicity profile, we construct {\sc APEC} spectra in {\tt Xspec} and obtain deprojected electron density profiles of the 1178 galaxy groups in our sample \citep[similarly done in][]{Liu2022} using the scaled temperature profiles and three metallicity profiles (low-scattered, median, and up-scattered $Z_{M17}$ profiles) shown with red in Fig.~\ref{fig:metallicity_prof}. We present the ratios of the electron densities in Fig.~\ref{fig:ne_ratio_t} that are obtained by using the aforementioned three metallicity profiles (low-scattered $Z_{M17}$: $n_{{\rm e},L}$, median $Z_{M17}$: $n_{{\rm e}}$, and up-scattered $Z_{M17}$: $n_{{\rm e},U}$) for 1178 groups at the three characteristic radii ($0.15r_{500c}$, $r_{2500c}$ and $r_{500c}$). The ratios ($n_{{\rm e},U}/n_{{\rm e}}$ and $n_{{\rm e},L}/n_{{\rm e}}$) ranging between $0.84 - 1.24$ in Fig.~\ref{fig:ne_ratio_t} indicating a non-negligible difference between the electron density measurements. The ratios deviate more from unity as the characteristic temperature decreases, and the measurement radius increases. This is due to line emission,  coupled with metalicity, which plays a more important role as the temperature decreases. The procedure described above is followed for the final results, and the radius/temperature dependent systematics due to the choice of metallicity profile are quantified and propagated to our final entropy measurements presented in Sect.~\ref{sec:results}. 

\subsection{Correction for the flux discrepancy}
\label{sec:missing_flux}

A discrepancy of $15 \%$ is reported in the luminosity measurements in the soft-band ($0.5-2$~keV) of a subsample of massive galaxy clusters observed with both \rosi\ and \chandra \citep{Bulbul2024}. The observed flux difference is constant with no flux or luminosity dependence. Some of this difference can be explained by the photon loss in the latest processing due to the higher CCD thresholds \citep{Merloni2024}; however, further investigation is required to understand the observed flux discrepancy, which could be due to various calibration effects. We account for the flux discrepancy while comparing our entropy measurements with those reported in the literature. Among our two main X-ray observables $n_{\rm e}$ and $T$, only electron density is impacted by the flux discrepancy since the spectroscopic $T$ measurements are not sensitive to the overall flux normalization. To roughly estimate the impact, we assumed the shape of the measured electron density profile to be the same for different instruments, used the fact that $L \propto n_{\rm e}^2$, and obtained a ratio of $n_{e,\rm eRO}/n_{e,\rm Cha} = 0.85^{0.5} \sim 0.92$ between the electron density measurements of \chandra and \rosi. The $8\%$ underestimation of $n_{\rm e}$ corresponds to a $5\%$ overestimation of entropy. This fraction is factored in the \chandra measurements in S09 when comparing with the \rosi\ results in Fig.~\ref{fig:Sall_T_lit_comp}.

\subsection{Systematics related to mass measurements}
\label{sec:sys_r500}
\begin{table*}
\centering
\caption{Summary of assumptions, corrections, and systematics.}
\label{tab:table_of_systematics}
\begin{tabular}{p{0.2\textwidth}p{0.75\textwidth}}
\toprule\midrule
Assumptions, corrections and systematics & Description and our approach$^\ast$ \\
\midrule
Temperature and metallicity profiles & Radial variation of temperature and metallicity need to be accounted for to have reliable thermodynamic profiles of galaxy groups. Given the limited signal-to-noise we have for most of the groups in our sample, we adopt the average shape of the temperature profiles of the galaxy groups presented in S09 and allow its normalization to vary for deriving the average thermodynamic properties of our sample. Furthermore, we adopt the average \citet{Mernier2017} metallicity profile for the main results and conservatively consider the reported scatter as the systematic uncertainty of the profile. We then propagate the systematic uncertainty to our final results and consider its impact as part of the total error budget.\\
Instrumental calibration & A flux mismatch of 15\% is reported in \citet{Bulbul2024} between \rosi and \chandra for galaxy clusters and groups. Assuming the flux mismatch is not a function of radii, this discrepancy corresponds to an 8\% difference in $n_{e}$ and a 5\% difference in $S$. In this work, we take the mismatch into account while comparing our results with the measurements in the literature with other telescopes. \\
Mass measurements & Obtaining the underlying mass distributions of galaxy groups are challenging and may lead to inconsistencies while comparing measurements. In this work, we account for the mass measurement mismatches while comparing our results with the literature and provide derivatives of our entropy measurements for future work to account for the mass measurement systematic while comparing with our results. \\
Atomic databases & Spectral models evolve over time as our knowledge of atomic transitions increases. This may result in discrepancies when measurements obtained with different atomic database versions are compared. In this work, we account for this by applying corrections to the measurements in the literature. \\

\bottomrule
\end{tabular}
\tablefoot{\\
$^\ast$ See Sect.~\ref{sec:systematics} for a more detailed description of the assumptions, corrections, and systematics, along with a more detailed prescription on how they are addressed in this study. \\
}
\end{table*}

We note that entropy measurements at overdensity radii, $0.15r_{500c}$, $r_{2500c}$, and $r_{500c}$ are sensitive to the assumed masses of the galaxy groups. This dependence is due to the entropy profile of galaxy groups being a strong function of the radial distance and the overdensity radius, which is a mass-dependent quantity \citep[e.g.][]{Bulbul2010, Ghirardini2019}. Therefore, any disagreement in radius and mass may lead to a bias in the measured thermodynamic profiles and their comparisons between different methods. Masses of galaxy groups can be estimated in different ways, such as by assuming hydrostatic equilibrium, using the shear information of the lensed galaxies, or using scaling relations; however, these methods have advantages and disadvantages along with introduced biases. Comparison of the mass estimation techniques for galaxy groups is beyond the scope of this paper; therefore, in this paper, we account for the bias introduced while comparing our results with the literature. We find that our scaling relations based $r_{500c}$ estimates are $\sim11\%$ higher than the hydrostatic equilibrium based $r_{500c}$ estimates in S09 for 9 crossmatched groups. Comparison between our characteristic radii estimates ($r_{500c}$) and those in the literature ($r_{500c,\rm S09}$) is provided in Fig.~\ref{fig:r500_comp}. A discrepancy of $\sim11\%$ approximately corresponds to a bias of $\sim37\%$ ($1.11^3 = 1.37$) on $M_{500c}$. This result is close but slightly below the $45\%$ hydrostatic mass bias observed in galaxy groups \citep{Nagai2007}, see also Sect.~6.2 of S09. We note that the masses of S09 are obtained with an outdated version of {\sc AtomDB}. \citet{Lovisari2015} and \citet{Sun2012} independently confirmed that using a more recent {\sc AtomDB} version (v2.0.1) increases the temperatures by $\sim15\%$. Such an increase would reduce the mass mismatch to a $\sim22\%$ level and the radius mismatch to $\sim7\%$ level.

One should note that constraining hydrostatic mass bias, especially in galaxy groups, is challenging due to our limited knowledge of the magnitude of non-thermal pressure support, magnified in galaxy groups due to powerful AGN feedback compared to galaxy clusters. Moreover, the fraction of discrepancy may be due to other systematics in the measurements, such as the representation of the $L_{X}-M$ relation with a single power-law spanning a large mass range. We provide the $r_{500c}$ estimates of the 9 crossmatched groups in Table~\ref{tab:r_500} along with the slopes of the average entropy profiles of the binned groups at the three characteristic radii in Table~\ref{tab:t_k_measurements}. 

The mass estimate dependence is also responsible for the observed scatter in the entropy of the sample. Fig.~\ref{fig:Sall_T_cc} shows a strong correlation between the scatter of the average entropy measurements at a given temperature and the average concentration ($c_{{\rm SB},r_{500c}} = SB(r<0.1r_{500c})/SB(r<r_{500c})$) of the sample (see Sanders et al., in prep. for the concentration measurements). This is expected since the $M_{500c}$ (or $r_{500c}$) estimates used in this work are obtained using an $L_{X}-M$ scaling relation. At a given mass, the cool-core galaxy groups with higher luminosity would have higher entropy measurements while the others scatter around the sample's median. The average entropy and temperature measurements of the sample are presented in Fig.~\ref{fig:Sall_T_cc}, where the colors of the data points indicating average concentration obtained by averaging the measurements provided in Sanders et al. (in prep.). A few extreme cases with a larger concentration, electron number density, and entropy are easily noticeable in Fig.~\ref{fig:Sall_T_cc}. By construction, the $r_{500c}$ (or $M_{500c}$) estimates are, on average, unbiased at the sample scale. Therefore, a small scatter does not significantly impact the conclusions in this work. On average, these effects cancel out such that at all three radii, the error-weighted average entropy plotted in magenta coincides with the intermediate concentration values $\log_{10}(c_{{\rm SB},r_{500c}}) \sim -0.7$.

\subsection{Systematics related to the atomic databases}
\label{sec:sys_atomic_database}
Over time, our knowledge of the strength of atomic transitions has changed significantly for $T<2~{\rm keV}$ plasma. The change in our knowledge of the strength and the width of the Fe-L complex is the most relevant to this work, given its key role in determining the temperature and metallicity of the IGrM. The change in our knowledge resulted in significant updates on the predictions of the spectral models, {\sc APEC} \citep[][]{Smith2001} and {\sc CIE} \citep[an updated version of the {\sc MEKAL} model,][]{Mewe1985,Mewe1986,Liedahl1995}, that retrieve data from atomic databases, {\tt AtomDB} \citep{2012Foster} and {\tt SPEXACT} \citep{Kaastra1996,Kaastra2020}, around the Fe-L complex \citep[e.g., see Fig. 1 in][]{Gastaldello2021}.

\begin{figure*}
\includegraphics[width=0.3\textwidth]{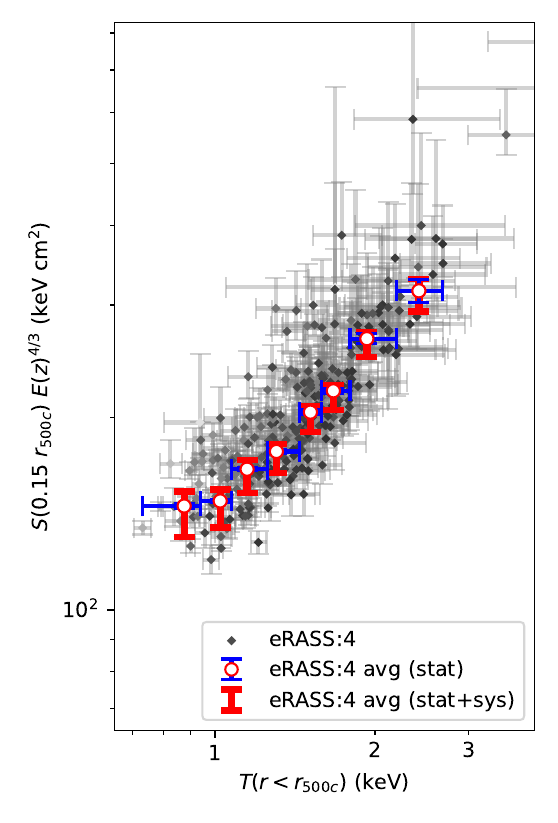} 
\includegraphics[width=0.303\textwidth]{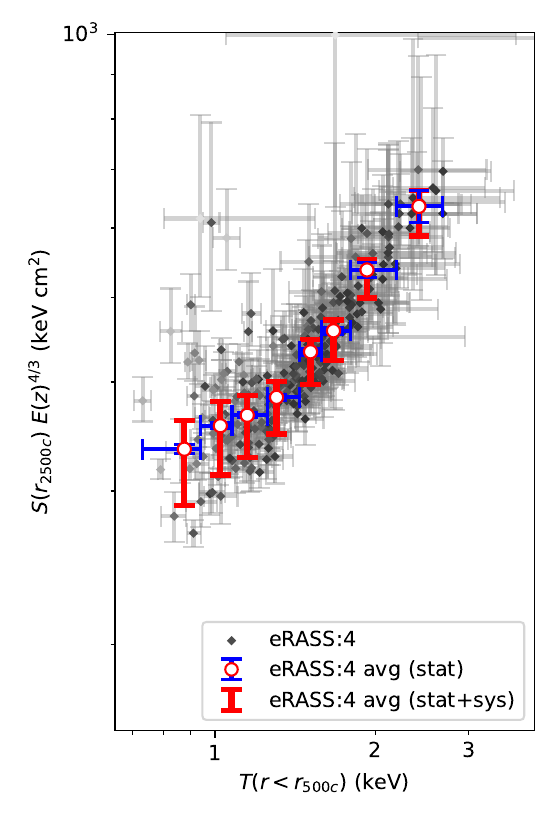} 
\includegraphics[width=0.376\textwidth]{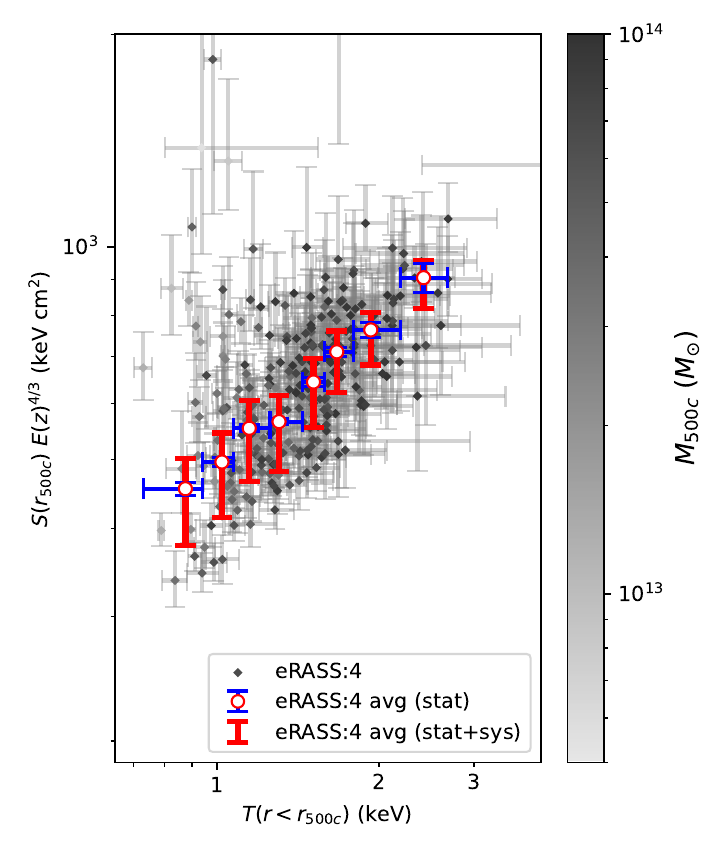}
\caption{Redshift evolution scaled average entropy measurements of binned groups (gray diamonds) at three radii ($0.15r_{500c}$, $r_{2500c}$, $r_{500c}$) as a function of characteristic temperature, $T(r<r_{500c})$. The colors of the gray data points represent the average masses ($M_{500c}$) of the groups within the corresponding Voronoi bin. Error-weighted averages of the data points are plotted as white circles, statistical uncertainties of the averages are shown with blue error bars, and the overall error budget of the average measurements resulting from the statistical and systematic uncertainties are shown with red error bars (see Sect.~\ref{sec:systematics} for the details of the accounted systematics).\label{fig:Sall_T_avg}}
\end{figure*}

These updates may cause measurement mismatches between the studies that employed different versions of the atomic databases. For this reason, the possible impact of using different versions of the atomic databases should be taken into account while comparing results with the literature. In fact, \citet{Sun2012} reports that for the $T<2~{\rm keV}$ plasma, the temperature measurements presented in S09 increase by  $10 - 20\%$ when a more recent {\tt AtomDB} version, v2.0.1, is used instead of the one employed in S09, v1.3.1. The increase in the temperature is also independently confirmed by \citet{Lovisari2015}, who found their temperature measurements to be 13\% higher than the temperatures of S09 for the seven crossmatched objects. In this work, we take this change into account and apply a 15\% correction to the entropy and temperature measurements of S09 to their $T<2~{\rm keV}$ groups while comparing their results with the eROSITA measurements (e.g., in Fig.~\ref{fig:Sall_T_lit_comp}). With this correction, we aim to have the fairest comparison of our results.

\citet{Lovisari2015} further report that in the case of such an update in the atomic database, the normalization of the {\sc APEC} spectrum reduces $\sim10\%$  for a group (NGC3402) residing at the low-T parameter space ($T\sim1$~keV), where the impact of the change is expected to be the most prominent. The density of the plasma scales as the square root of the {\sc APEC} normalization, which results in a $\sim3\%$ increase in entropy. Given that the $3\%$ increase in entropy is an upper limit \citep[as explained in section A.3 of][]{Lovisari2015} and is well within the entropy error bars presented in S09, we did not propagate the impact of this change to the S09 entropy measurements.

Lastly, \citet{Sun2012} reported that abundance measurements drop $\sim20\%$ after updating the {\tt AtomDB} version (from 1.3.1 to 2.0.1). This reduces the stacked metallicity data points of \citet{Sun2012} by $\sim20\%$; however, given that we are employing \citet{Mernier2017} profile in our work, our results are not affected by this. We further note that even if we apply such a correction to \citet{Sun2012} measurements, they will be well within our conservative systematic error bars for the metallicity profile (shaded area in Fig.~\ref{fig:metallicity_prof}); therefore, it does not pose any challenge to our measurements.

It can be seen from Fig.~1 of \citet[][]{Gastaldello2021} that the width and the normalization of the Fe-L complex seemed to be converging. However, it is hard to know how far we are from the absolute calibration. Therefore, future work should keep systematics related to the atomic databases in mind while comparing results from the literature.

\section{Entropy and characteristic temperature measurements}
\label{sec:results}

In this work, we constrain the entropy of the IGrM, utilizing the deep eRASS:4 observations of the galaxy groups detected in the eRASS1 survey. In this section, we present our measurements of the characteristic temperature, $T(r<r_{500c})$, and entropy at three overdensity radii, $0.15r_{500c}$, $r_{2500c}$, and $r_{500c}$, from the combined analysis of 1178 galaxy groups. We then compare our findings with the previously reported results in the literature.

Following the procedures described in Sects.~\ref{sec:sample_selection} and \ref{sec:data_analysis}, we obtain average entropy and temperature measurements of a sample of 1178 \rosi\ selected galaxy groups in 271 bins. Our measurements for the average temperature and average entropy at three characteristic radii scaled by the self-similar redshift evolution ($E(z)^{4/3}$) can be seen in Fig.~\ref{fig:Sall_T_avg}, where the error-weighted average entropy measurements are shown in white circles, the statistical uncertainties are shown with blue error bars and the overall error budget resulting from the statistical and systematic (see Sect.~\ref{sec:systematics}) uncertainties are shown with red error bars. Error-weighted average of the redshift evolution scaled entropy measurements, their statistical uncertainties, and the impact of systematics are also provided in Table~\ref{tab:t_k_measurements} as a function of IGrM temperature. Besides the average entropy measurements, we provide slopes of the average entropy profiles in Table~\ref{tab:t_k_measurements} at the three characteristic radii for future work to account for the mass measurement systematic while comparing with our results (see Sect.~\ref{sec:sys_r500} for the details of accounting the systematics in mass measurements). We note that the few outlying entropy measurements in Fig.~\ref{fig:Sall_T_avg} are due to the extreme cool-core objects and have negligible impact on our error-weighted entropy measurements (see Sect.~\ref{sec:sys_r500} for a more detailed discussion on the outliers and their impact on our final results).

\begin{table*}
\centering
\caption{Average entropy and entropy slope measurements of the grouped sample as a function of temperature at the three characteristic radii $0.15r_{500c}$, $r_{2500c}$, and $r_{500c}$.}
\label{tab:t_k_measurements}
\resizebox{\textwidth}{!}{\begin{tabular}{lcccccccc}
\toprule\midrule
\multicolumn{1}{c}{}       & \multicolumn{8}{c}{$T(r<r_{500c})$ (keV)} \\
\cmidrule(lr){2-9}
& $0.73-0.94$ & $0.94-1.08$ & $1.08-1.26$ & $1.26-1.44$ & $1.44-1.59$ & $1.59-1.79$ & $1.79-2.19$ & $2.19-2.68$  \\
\midrule
$S(0.15r_{500c})E(z)^{4/3}$$^{\ast}$ & ${145.4}_{-1.1(-15.3)}^{+1.1(+7.7)}$ & ${148.09}_{-0.89(-13.5)}^{+0.90(+6.1)}$ & ${166.04}_{-0.86(-13.6)}^{+0.87(+5.5)}$ & ${177.0}_{-1.2(-13.2)}^{+1.2(+4.6)}$ & ${203.9}_{-1.6(-14.0)}^{+1.6(+4.3)}$ & ${220.2}_{-2.0(-14.4)}^{+2.1(+3.9)}$ & ${265.8}_{-4.9(-16.5)}^{+5.0(+3.9)}$ & ${316}_{-13(-18.6)}^{+13(+3.6)}$ \\
$S(r_{2500c})E(z)^{4/3}$$^{\ast}$ & ${335.0}_{-4.1(-46)}^{+4.1(+26)}$ & ${356.3}_{-2.9(-43)}^{+2.9(+24)}$ & ${366.3}_{-2.3(-39)}^{+2.3(+20)}$ & ${384.0}_{-2.6(-35)}^{+2.7(+16)}$ & ${433.0}_{-3.7(-35)}^{+3.7(+14)}$ & ${457.7}_{-4.4(-34)}^{+4.5(+12)}$ & ${537}_{-10(-37)}^{+10(+11)}$ & ${635}_{-26(-40.0)}^{+27(+9.8)}$ \\
$S(r_{500c})E(z)^{4/3}$$^{\ast}$ & ${454.9}_{-9.4(-77)}^{+9.6(+46)}$ & ${496.3}_{-7.6(-81)}^{+7.7(+49)}$ & ${553.9}_{-6.1(-88)}^{+6.1(+52)}$ & ${566.0}_{-5.9(-84)}^{+5.9(+50)}$ & ${644.0}_{-8.6(-88)}^{+8.7(+51)}$ & ${710.6}_{-9.9(-88)}^{+10.1(+49)}$ & ${763}_{-18(-80)}^{+18(+41)}$ & ${905}_{-42(-76)}^{+44(+32)}$ \\
\midrule
$S'(0.15r_{500c})^\dagger$ & $0.76$ & $0.82$ & $0.67$ & $0.71$ & $0.63$ & $0.60$ & $0.61$ & $0.61$ \\
$S'(r_{2500c})^\dagger$ & $0.61$ & $0.64$ & $0.68$ & $0.63$ & $0.65$ & $0.65$ & $0.59$ & $0.59$ \\
$S'(r_{500c})^\dagger$ & $0.30$ & $0.23$ & $0.37$ & $0.34$ & $0.46$ & $0.59$ & $0.44$ & $0.41$ \\
\bottomrule
\end{tabular}}
\tablefoot{\\
$^\ast$ Error-weighted average of the redshift evolution scaled entropy measurements of the grouped sample in units of ${\rm keV~cm^2}$ within the corresponding temperature bin. The first set of errors above and below the measurements represent the statistical uncertainty, and the errors presented within the parenthesis represent the systematic uncertainty (see Sect.~\ref{sec:systematics} for the details of the accounted systematics).  \\
$^\dagger$ $S'(r) = \frac{d\log(S(r))}{d\log(r/r_{500c})}$, the slope of the error-weighted average entropy measurements as a function of dimensionless radius in logarithmic space, in units of ${\log({\rm keV~cm^2})}$. \\
}
\end{table*}

We find that the characteristic temperature measurements of our binned group sample span a range of $\sim0.73 - 2.68$ keV. Furthermore, we find that the characteristic temperature measurements of the binned groups correlate well with the entropy measurements at the three characteristic radii such that, an increase in the characteristic temperature from $0.73$~keV to $2.68$~keV corresponds to an increase from the redshift scaled entropy levels of $121$, $271$, $341~{\rm keV~cm^2}$ to $404$, $722$, $1135~{\rm keV~cm^2}$ respectively. This trend can also be clearly seen from the error-weighted average measurements (white circles) in Fig.~\ref{fig:Sall_T_avg} such that the error-weighted averages at these three radii increase from the levels of $145$, $335$, $455~{\rm keV~cm^2}$ to $316$, $635$, $905~{\rm keV~cm^2}$ respectively. It can further be noticed that as the characteristic temperature of the binned groups decreases, the statistical uncertainty of the error-weighted average profiles (blue error bars) decreases. Conversely, the decrease in the temperature corresponds to an increase in the overall error budget (red error bars). This is due to the fact that the line emission from the Fe-L complex at the low-temperature parameter space becomes significant, which provides an additional spectral feature for measuring the temperature and reduces the statistical uncertainty. Meanwhile, at the same parameter space, the systematic uncertainty of the metallicity profile has a very large impact on the electron density measurements, which results in the entropy measurements having large systematic uncertainties. At higher temperatures, the systematic uncertainties of the electron density measurements become smaller due to the reduced line emission, and the less significant Fe-L complex increases the statistical uncertainties.

\begin{figure*}
\includegraphics[width=0.99\textwidth]{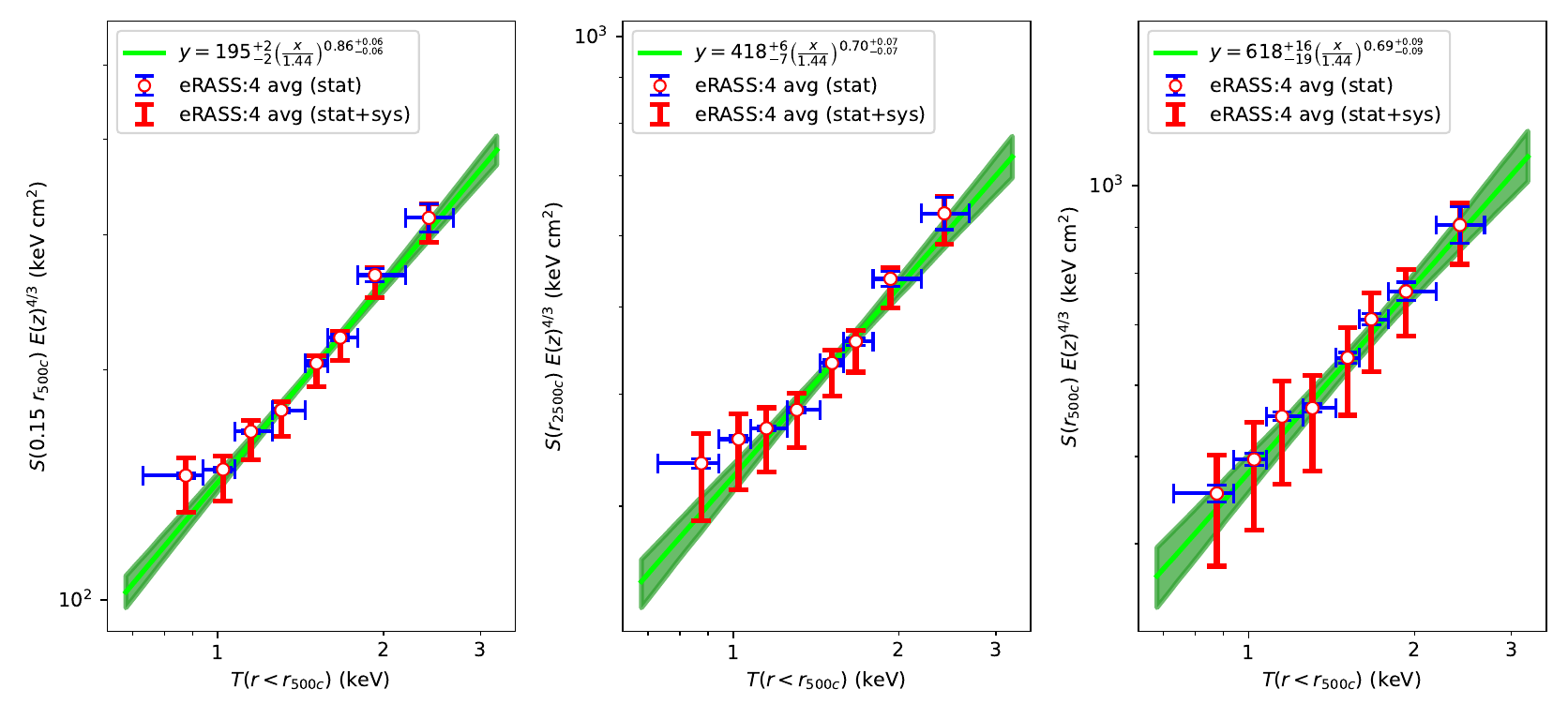}
\caption{$S(r)-T$ relations at $0.15r_{500c}$, $r_{2500c}$ and $r_{500c}$. The white circles represent redshift evolution scaled average entropy measurements, blue error bars represent statistical uncertainties of the averages, and the red error bars represent the overall error budget of the average measurements resulting from the statistical and systematic uncertainties (see Sect.~\ref{sec:systematics} for the details of the accounted systematics). The best-fit power-law models to the data points are plotted as light green lines, and the uncertainties of the best-fit lines are shown with dark green shaded regions.\label{fig:Sall_T_fit}}
\end{figure*}

Furthermore, we find that the redshift scaled average entropy and the average temperature measurements of our sample at the three radii follow power-law relations within the uncertainties. To quantify the normalizations and the slopes of the relations, we fit a power-law model to the measurements of the form $S(r) E(z)^{4/3} = A(T/T_{\rm piv})^{B}$ where the $S(r)$ term stands for entropy measured at three characteristic radii ($0.15r_{500c}$, $r_{2500c}$, and $r_{500c}$), the $T$ term stands for the characteristic temperature ($T(r<r_{500c})$), $T_{\rm piv}$ term stands for the pivot temperature, and the $A$ and $B$ terms stand for the normalization and the slope of the relation respectively. For our galaxy group sample, we took a pivot value of $T_{\rm piv}=1.44~\rm keV$ and obtained the best-fit power-law models to the average measurements at the three radii as shown in Fig.~\ref{fig:Sall_T_fit}. Furthermore, we noticed that the slope of the $S(r)-T$ relations seem to be changing around $T = 1.44~\rm keV$ such that the warm/higher-mass groups in our sample ($T = 1.44-2.68~\rm keV$) are steeper than the $S(r)-T$ relations of the cool/lower-mass groups ($T = 0.73-1.44~\rm keV$). To quantify the difference, we separately fit the average measurements of the warm and cool groups using the same relation, assuming pivot values of $T_{\rm piv}=1.79$ and $1.08~\rm keV$, respectively. Best-fit parameters of the $S(r)-T$ relation at three radii for all galaxy groups in our sample, along with the best-fit parameters for the warm, and cool groups, are listed in Table~\ref{tab:slopes}. 

As a result of the fitting procedure, we find that for all galaxy groups ($T = 0.73-2.68~\rm keV$), the slopes of the $S(r_{2500c})-T(r<r_{500c})$ and $S(r_{500c})-T(r<r_{500c})$ relations ($B=0.70^{+0.07}_{-0.07}$ and $0.69^{+0.09}_{-0.09}$ respectively) are in very good agreement with each other whereas we find that the slope of the $S(0.15r_{500c})-T(r<r_{500c})$ relation ($B=0.86^{+0.06}_{-0.06}$) being slightly steeper. We also compare our relations with the self-similar predictions presented in \citet{Voit2005}. We find that our best-fit normalizations of the $S(r)-T$ relations are $>5\sigma$ larger than the self-similar predictions at all radii. This agrees well with the previous findings on the entropy excess in galaxy groups and indicates that non-gravitational processes such as AGN feedback and radiative cooling are non-negligible and play an important role in shaping the entropy profiles of galaxy groups. Moreover, we compared the best-fit slopes of the $S(0.15r_{r500c})-T(r<r_{500c})$, $S(r_{2500c})-T(r<r_{500c})$, $S(r_{500c})-T(r<r_{500c})$ relations for all galaxy groups in our sample with the self-similar prediction ($B_{\rm self}=1$ at all radii) and found that our slopes are $2.3$, $4.3$, and $3.4\sigma$ shallower than the self-similar slope. 

Furthermore, we compared the best-fit $S(r)-T$ relations of the warm and cool groups and found that the slopes of the warm/higher-mass groups, $B=0.98^{+0.10}_{-0.10}$, $0.87^{+0.11}_{-0.11}$ and $0.72^{+0.17}_{-0.17}$, are much steeper than the slopes of the cool/lower-mass groups, $B=0.59^{+0.16}_{-0.16}$, $0.32^{+0.22}_{-0.22}$ and $0.56^{+0.28}_{-0.28}$ at $0.15r_{500c}$, $r_{2500c}$ and $r_{500c}$ respectively. We then compared our best-fit normalizations of the warm and cool groups with the self-similar predictions presented in \citet{Voit2005} and found that the normalizations for warm and cool groups are $>5\sigma$ larger than the self-similar predictions at all radii. Moreover, we also compared the slopes of the warm and cool groups with the self-similar prediction and found that the best-fit slopes of the warm/higher-mass groups, $B=0.98^{+0.10}_{-0.10}$, $0.87^{+0.11}_{-0.11}$ and $0.72^{+0.17}_{-0.17}$, agree much better with the self-similar prediction ($B_{\rm self} = 1$) compared to the slopes of the cool/lower-mass groups, $B=0.59^{+0.16}_{-0.16}$, $0.32^{+0.22}_{-0.22}$ and $0.56^{+0.28}_{-0.28}$, at $0.15r_{500c}$, $r_{2500c}$, and $r_{500c}$ respectively. This comparison suggests that even though the non-gravitational processes significantly increase the overall entropy levels of galaxy groups at all temperature/mass scales, they result in the slope of the $S(r)-T$ relation for cool/lower-mass groups deviating more from the self-similar prediction compared to the slope of the warm/higher-mass groups that live at a temperature/mass parameter space closer to clusters. 

We note that the observed flattening for the cool groups could also be due to other reasons, such as the selection effects or relatively large systematic uncertainties. The X-ray selection probability of galaxy clusters and groups has a dependency on the emission profile of the cluster and group \citep{Clerc2024}. Therefore, for the most robust calibration of the $S(r)-T$ scaling relation, a selection function that takes entropy at given radii as input should be used, which is currently not implemented in our framework for galaxy groups. In the future, a better understanding of the temperature and metallicity profiles of galaxy groups, along with the use of a profile-dependent selection function for fitting the $S(r)-T$ relation, would make the picture clearer and help us understand the origin of the observed flattening for the cool/lower-mass groups.

Comparing the entropy measurements at the three radii, with the previous measurements in the literature as a function of temperature, $T(r<r_{500c})$, is challenging given the limited number of studies reporting these quantities and the information provided in these studies to account for the discrepancies in the measurement radii being limited across the literature. In this section, we present our findings from the comparison performed between our measurements and the results reported by S09 and \citet{Johnson2009}.

\begin{table}
\centering
\caption{Best-fit parameters of the $S(r)-T$ relation.}
\label{tab:slopes}
\begin{tabular}{lcc}
\toprule\midrule
Relation   & $A$  & $B$ \\
 & $(\rm keV~cm^2)$  & \\
\midrule
\multicolumn{3}{c}{{\bf All groups} ($T = 0.73-2.68~\rm keV$, $T_{\rm piv} = 1.44~\rm keV$)} \\
\midrule
$S(0.15r_{500c})-T(r<r_{500c})$ & $195^{+2}_{-2}$ & $0.86^{+0.06}_{-0.06}$ \\
$S(r_{2500c})-T(r<r_{500c})$ & $418^{+6}_{-7}$ & $0.70^{+0.07}_{-0.07}$ \\
$S(r_{500c})-T(r<r_{500c})$ & $618^{+16}_{-19}$ & $0.69^{+0.09}_{-0.09}$ \\
\midrule
\multicolumn{3}{c}{{\bf Warm groups} ($T = 1.44-2.68~\rm keV$, $T_{\rm piv} = 1.79~\rm keV$)} \\
\midrule
$S(0.15r_{500c})-T(r<r_{500c})$ & $235^{+4}_{-4}$ & $0.98^{+0.10}_{-0.10}$ \\
$S(r_{2500c})-T(r<r_{500c})$ & $485^{+9}_{-10}$ & $0.87^{+0.11}_{-0.11}$ \\
$S(r_{500c})-T(r<r_{500c})$ & $717^{+24}_{-26}$ & $0.72^{+0.17}_{-0.17}$ \\
\midrule
\multicolumn{3}{c}{{\bf Cool groups} ($T = 0.73-1.44~\rm keV$, $T_{\rm piv} = 1.08~\rm keV$)} \\
\midrule
$S(0.15r_{500c})-T(r<r_{500c})$ & $155^{+3}_{-4}$ & $0.59^{+0.16}_{-0.16}$ \\
$S(r_{2500c})-T(r<r_{500c})$ & $352^{+11}_{-13}$ & $0.32^{+0.22}_{-0.22}$ \\
$S(r_{500c})-T(r<r_{500c})$ & $504^{+21}_{-27}$ & $0.56^{+0.28}_{-0.28}$ \\
\bottomrule
\end{tabular}
\tablefoot{The fitted relation is of the form $S(r) E(z)^{4/3} = A(T/T_{\rm piv})^{B}$ where the $S(r)$ term stands for entropy measured at three characteristic radii  ($0.15r_{500c}$, $r_{2500c}$, and $r_{500c}$), the $T$ term stands for the characteristic temperature, $T(r<r_{500c})$, and $T_{\rm piv}$ is the pivot temperature value. The relation is fitted to the measurements presented in Table~\ref{tab:t_k_measurements} by taking into account the statistical and systematic uncertainties. The fitting procedure is executed three times at each radius: first, for all the groups in the sample spanning a temperature range of $0.73-2.68~\rm keV$; second, for only the warm groups spanning a temperature range of $1.44-2.68~\rm keV$; and third, for only the cool groups spanning a temperature range of $0.73-1.44~\rm keV$.}
\end{table}

\begin{figure*}
\includegraphics[width=0.99\textwidth]{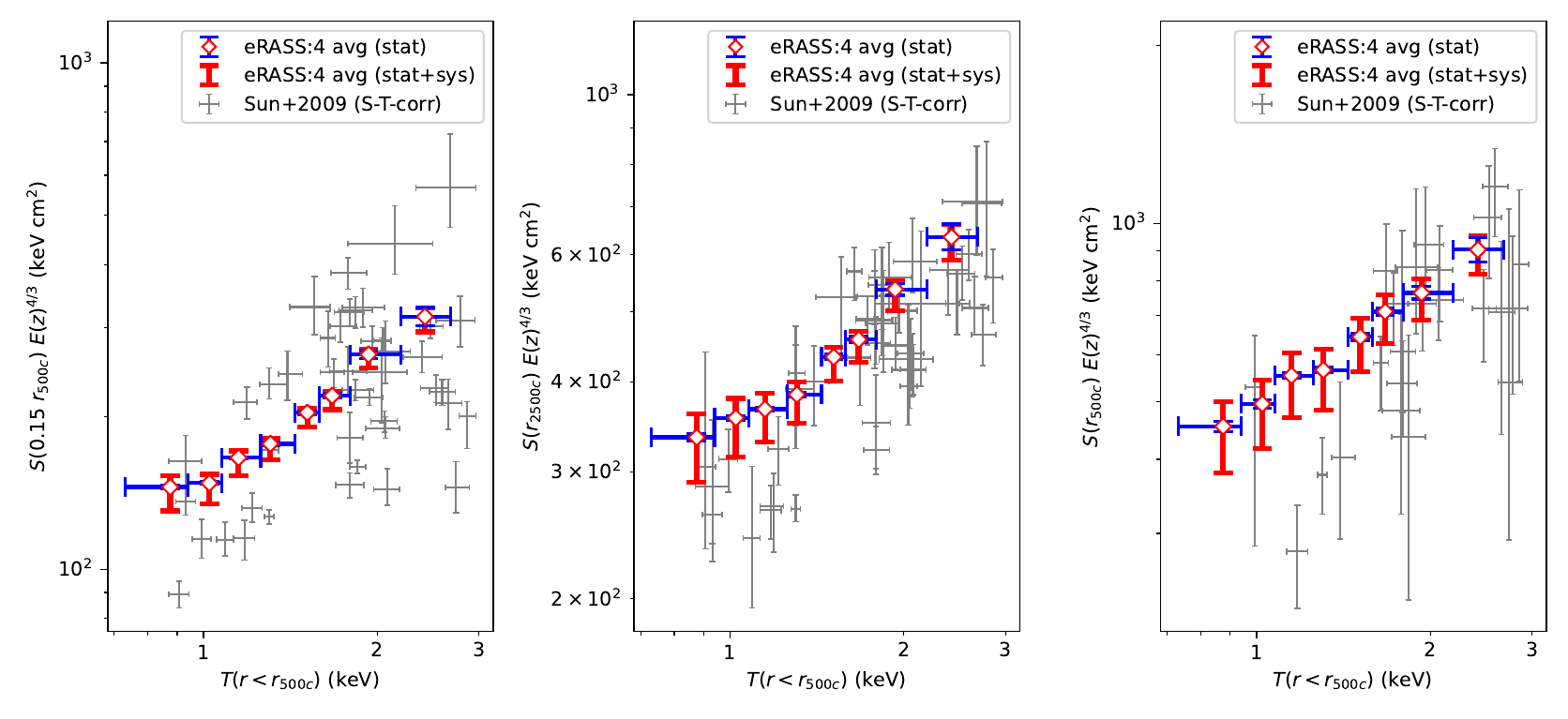}
\caption{Comparison between the redshift evolution scaled average eROSITA entropy measurements (white diamonds) with the \chandra measurements of 43 galaxy groups (black error bars) presented in S09 at three radii ($0.15r_{500c}$, $r_{2500c}$, $r_{500c}$) as a function of characteristic temperature, $T(r<r_{500c})$. Blue error bars represent the statistical uncertainties of the average measurements, and the red error bars represent the overall error budget resulting from the statistical and systematic uncertainties (see Sect.~\ref{sec:systematics} for the details of the accounted systematics). For consistency, core-excised temperatures presented in S09 are converted to core-included temperatures, and the entropy measurements presented in S09 are normalized such that the new data points are measured at the same angular radii with \rosi\ flux calibration (see Sect.~\ref{sec:results}  for the details of these corrections). \label{fig:Sall_T_lit_comp}}
\end{figure*}
\begin{figure*}
\includegraphics[width=0.99\textwidth]{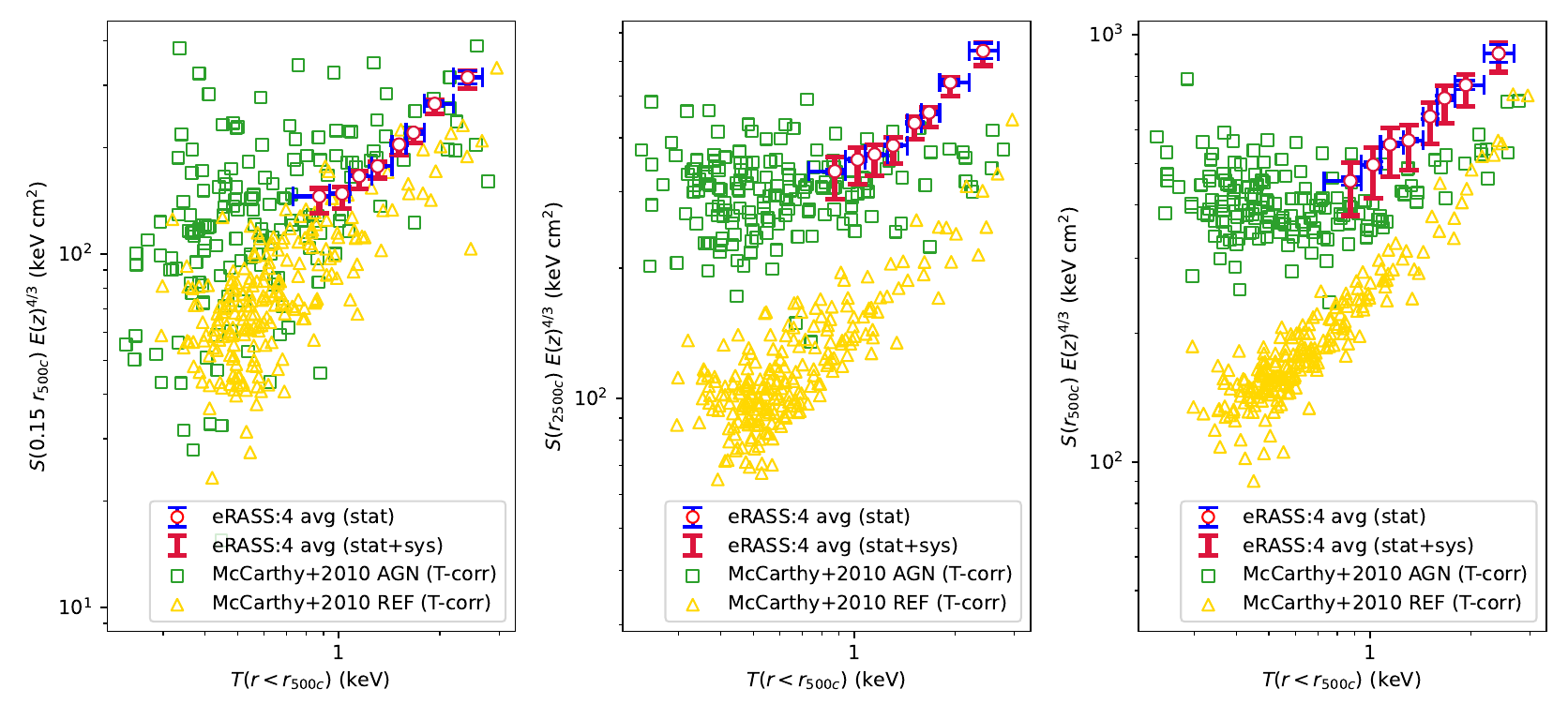} 
\caption{Comparison between the redshift evolution scaled average eROSITA entropy measurements (white circles) with the predictions of the REF (AGN feedback off) and AGN (AGN feedback on) runs of OWL simulations (yellow triangles and green squares) presented in \citet{McCarthy2010} at three radii ($0.15r_{500c}$, $r_{2500c}$, $r_{500c}$) as a function of characteristic temperature, $T(r<r_{500c})$. Blue error bars represent the statistical uncertainties of the average measurements, and the red error bars represent the overall error budget resulting from the statistical and systematic uncertainties (see Sect.~\ref{sec:systematics} for the details of the accounted systematics). For consistency, core-excised temperatures presented in \citet{McCarthy2010} are converted to core-included temperatures (see Sect.~\ref{sec:sim_comparison}  for the details of the correction).\label{fig:OWL_simulations_comp}}
\end{figure*}
\begin{figure*}
\centering
\includegraphics[width=0.49\textwidth]{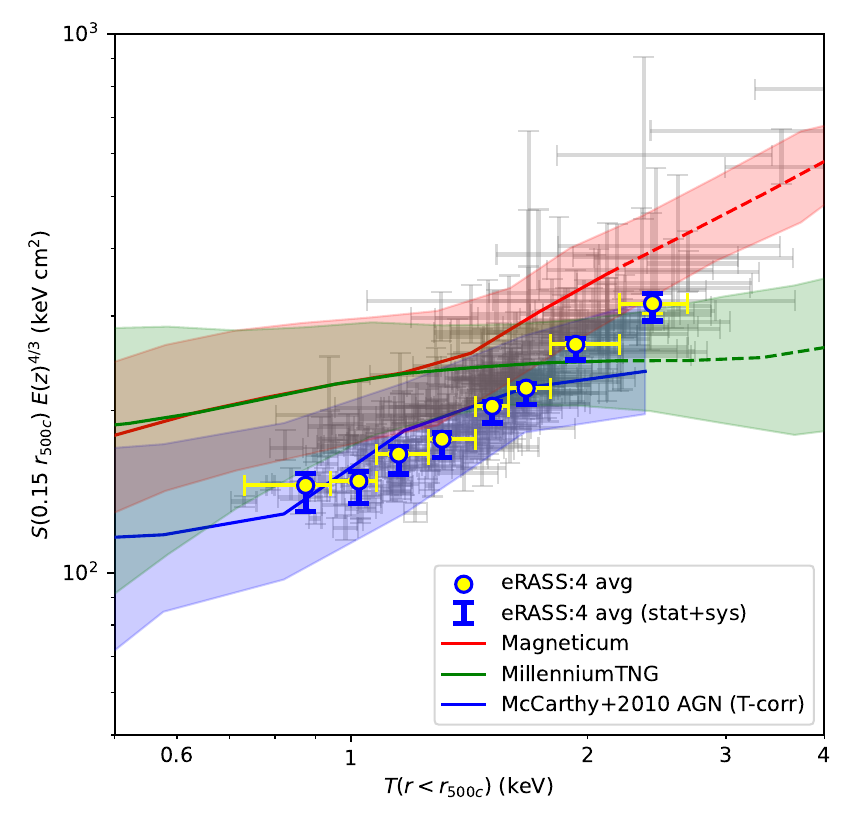} 
\includegraphics[width=0.49\textwidth]{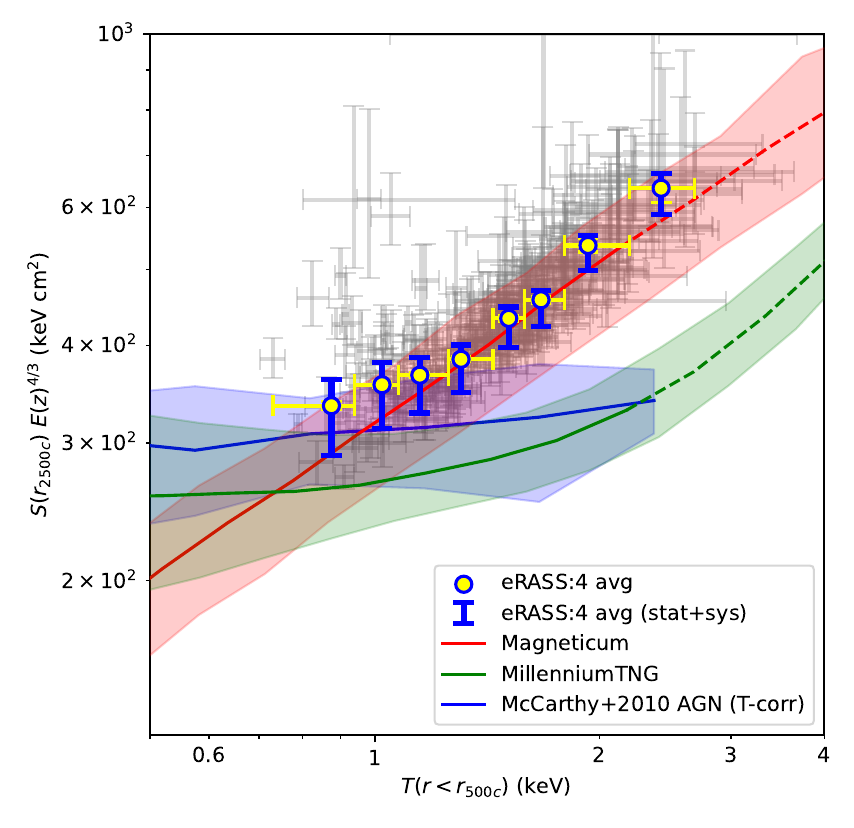} 
\includegraphics[width=0.49\textwidth]{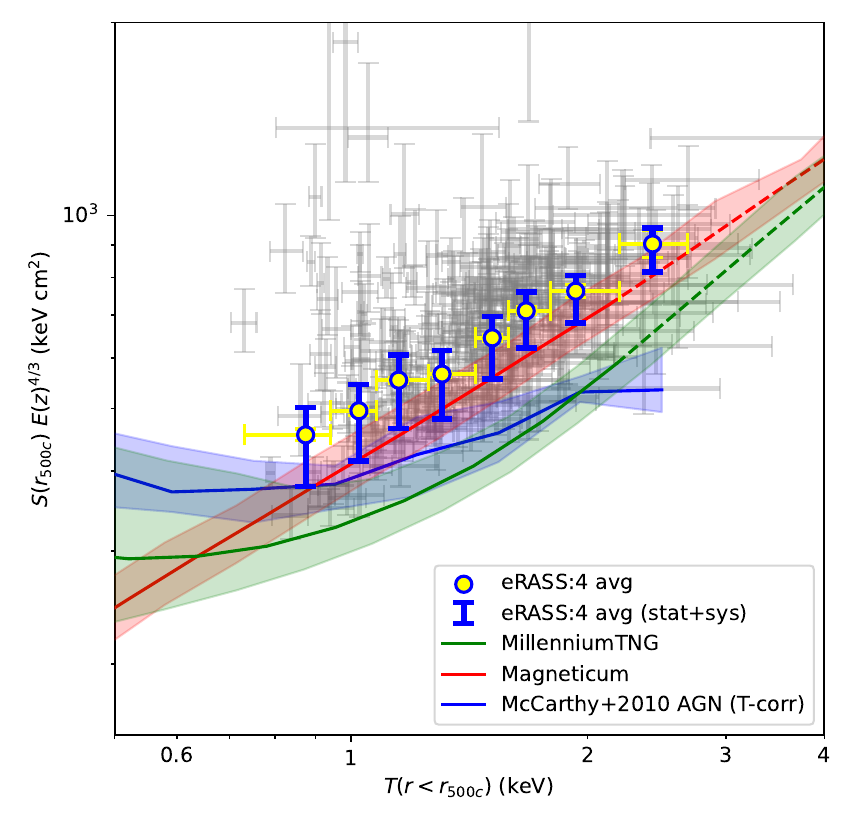} 

\caption{Comparison of the redshift evolution scaled average eROSITA entropy measurements (yellow circles) with the predictions of three simulations (Magneticum, MillenniumTNG, and OWL simulations) at $0.15r_{500c}$ (top left), $r_{2500c}$ (top right), and $r_{500c}$ (bottom) as a function of characteristic temperature, $T(r<r_{500c})$. Gray crosses represent the average entropy measurements of the binned groups, and the blue error bars represent the overall error budget of the average measurements resulting from the statistical and systematic uncertainties (see Sect.~\ref{sec:systematics} for the details of the accounted systematics). The solid lines represent the predictions of simulations at the group regime ($5\times10^{12}<M_{500c}<10^{14}~M_{\odot}$), the dashed lines represent the predictions at the low-mass cluster regime ($10^{14}<M_{500c}<3\times10^{14}~M_{\odot}$) and the shaded regions represent the scatter of the measurements. \label{fig:Simcomps_T}}
\end{figure*}

To have a fair comparison, we quantify and account for the systematic differences between our results and those presented in S09. The average temperature of galaxy groups in S09 are measured using core-excised apertures ($0.15r_{500c}<r<r_{500c}$), whereas in this work, we measure temperatures including the core out to $r_{500c}$ to maximize the signal-to-noise in our measurements. We convert the core-excised characteristic temperature measurements, $T(0.15r_{500c}<r<r_{500c})$, reported in S09 to the core-included temperatures by applying a conversion factor of $T(r<r_{500c})/T(0.15r_{500c}<r<r_{500c}) = 1.07$. This conversion factor is obtained by projecting the average group temperature profile (see Sect.~\ref{sec:sys_kT_metal}) using the temperature weighting and projection formulation of \citet{ZuHone2023} calibrated for \rosi\  and the average group electron density profile (Bahar et al., in prep.). Additionally, we apply a correction factor to account for the difference in the characteristic radii estimation to the S09 entropies as described in Sect.~\ref{sec:sys_r500}. The correction factors are obtained by comparing our scaling relations based on $r_{500c}$ estimates with the masses obtained assuming hydrostatic equilibrium (see Sect.~\ref{sec:sys_kT_metal}) and quantifying the impact of the mismatch on the average entropy measurements at three characteristic radii. Furthermore, we apply a 5\% correction to the entropy measurements of S09 to account for the 15\% flux discrepancy between \chandra and \rosi\ (see Sect.~\ref{sec:missing_flux}) that is reported in \citet{Bulbul2024}. Lastly, we applied a 15\% correction to the entropy and temperature measurements of the $T<2~{\rm keV}$ groups in the S09 sample due to the use of an old version of the atomic database {\tt AtomDB} (see Sect.~\ref{sec:sys_atomic_database} for the details on the correction).

The comparison between our temperature and entropy measurements with the findings of S09 are shown in Fig.~\ref{fig:Sall_T_lit_comp}. After the systematic differences are accounted for, we find our measurements to agree well with the S09 results within $\sim1\sigma$ confidence at all three radii. Fig.~\ref{fig:Sall_T_lit_comp} shows that our results only deviate from the measurements of S09 at the cooler temperatures with $T(r<r_{500c})<1.15$~keV. We argue that this may be due to the completeness of our sample being much higher than the completeness of the S09 sample at these temperatures. The S09 sample is based on the archival \chandra follow-up observations of 43 \rosat-detected bright groups. For this reason, by construction, the limited number of measurements reported in S09 at the low-temperature parameter space is for bright groups that have relatively high $n_{\rm e}$ and low $S$. This effect is visible in Fig.~\ref{fig:Sall_T_lit_comp} such that S09 has only three measurements for $S(r_{500c})$ and 11 for $S(r_{2500c})$ and $S(0.15r_{500c})$ at the cool temperatures end, whereas our sample includes temperature measurements of a sample of 323 galaxy groups binned into 61 groups. Furthermore, we note that the sample studies analyzing individually followed-up systems, such as S09, exclude morphologically disturbed systems. Nevertheless, in our work, with the aim of achieving comprehensive conclusions about galaxy groups, we did not make such a distinction and have analyzed all the galaxy groups in our sample by allowing the centroid of the surface brightness profile to be free. Parallel to having a preferential selection toward bright objects, excluding morphologically disturbed systems can also potentially decrease the average entropy of a sample at a given mass/temperature. That is in agreement with the direction of the slight discrepancy we see in Fig.~\ref{fig:Sall_T_lit_comp} and can possibly explain the mismatch. A more quantitative statement on the relationship between the morphology and gas properties of galaxy clusters and groups, as well as the relationship between the eROSITA selection and the morphology of the extended objects, will be explored in Sanders et al. (in prep.).

Furthermore, \citet{Johnson2009} analyzed \xmm observations of 28 nearby galaxy groups and found that the entropy measurements of their sample at the core, $S(0.1r_{500c})$, follow a power-law relation with their core-excluded temperature measurements, $T(0.1r_{500c}<r<0.3r_{500c})$. They further report that the best-fit slope of their $S(0.1r_{500c}) - T(0.1r_{500c}<r<0.3r_{500c})$ relation ($0.79 \pm 0.06$) agrees with the slope of a similar relation, $S(0.15r_{500c}) - T(0.15r_{500c}<r<r_{2500c})$, presented in S09, $0.78 \pm 0.12$. In our work, we measure entropy at the core at a slightly different radius ($0.15r_{500c}$) than the one used in \citet{Johnson2009}, and we use core-included temperature measurements rather than the core-excised temperatures. Nevertheless, we compare our most similar relation, $S(0.15r_{500c}) - T(r<r_{500c})$, to the relations fitted in S09 and \citet{Johnson2009} and find that our slope $0.86 \pm 0.08$ is in good statistical agreement with the slopes of S09 and \citet{Johnson2009}. Moreover, we further compare our best-fit slope for the $S(r_{2500c}) - T(r<r_{500c})$ relation with the best-fit slope of another similar relation, $S(r_{2500c}) - T(0.15r_{500c}<r<r_{500c})$, presented in S09 and find that our slope ($0.7 \pm 0.07$) agrees well with the slope reported in S09 ($0.76 \pm 0.06$). Lastly, we compare the best-fit slope of our $S(r_{500c}) - T(r<r_{500c})$ relation with the best-fit slope of S09 for the $S(r_{500c}) - T(0.15r_{500c}<r<r_{500c})$ relation and find that our slope ($0.69\pm0.09$) is in good statistical agreement with the slope reported in S09 ($0.8 \pm 0.2$).

\section{Comparison with the numerical simulations}
\label{sec:sim_comparison}

In this section, we compare our results with the state-of-the-art simulations to place constraints on the physics of the AGN feedback. We quantify the agreement or disagreement between our measurements and the predictions of the various AGN feedback models implemented in these cosmological hydrodynamical simulations, including MillenniumTNG\footnote{\url{https://www.mtng-project.org}} \citep{Hernandez2023,Pakmor2023}, Magneticum\footnote{\url{http://www.magneticum.org}} \citep{Hirschmann2014} and the OverWhelmingly Large Simulations \citep[OWL simulations,][]{Schaye2010}.

We compare our results with the Magneticum, MillenniumTNG, and OWL simulations that include different implementations for AGN feedback. The entropy measurements of the OWL simulations are plotted in Fig. 2 of \citet{McCarthy2010}; therefore, we extracted their measurements from their paper and directly used them in this work. In contrast, the entropy and the characteristic temperature profile measurements for MillenniumTNG and Magneticum simulations are not publicly available. For this reason, we extract thermodynamic profiles of the gaseous halos from the MillenniumTNG and Magneticum simulations. A brief description of the MillenniumTNG and Magneticum simulations and the extraction process is as follows.

\subsection{MillenniumTNG simulations}

The MTNG740 flagship full physics run of the MillenniumTNG project \citep{Pakmor2023} is used in this work, and it simulates a $500\,\mathrm{Mpc/h}$ cosmological box with the \citet{Planck2016} cosmology. Its galaxy formation model is close to the IllustrisTNG model \citep{Weinberger2017,Pillepich2018} and includes primordial and metal line cooling, a sub-grid model for star formation and the interstellar medium, mass return from stars via AGB stars and supernovae, an effective model for galactic winds, as well as a model for the formation, growth, and feedback from supermassive black holes. At a baryonic mass resolution of $3\times 10^7 \mathrm{M_\odot}$ MTNG740 reproduces well many properties of observed galaxies and galaxy clusters for halos with $M_\mathrm{500c}>2.3 \times 10^{12}\,\mathrm{M_\odot}$.

 The hydrodynamical profiles of groups shown in this work are computed the same way as the galaxy cluster profiles in \citet{Pakmor2023}. MTNG outputs a number of useful quantities: gas cell mass ($m$), electron abundance ($x$), and internal energy ($\epsilon$). Assuming a primordial hydrogen mass fraction of $X_H=0.76$, and an adiabatic index of $\gamma=5/3$, we compute the volume-weighted electron number density, $n_{\rm e}$, and temperature $T$, for each gas particle $i$ as

\begin{equation}
V_i n_{{\rm e},i} = x_i m_i \frac{X_H}{m_p}
\end{equation}
\begin{equation}
T_{i} = (\gamma-1) \epsilon_i/k \frac{4 m_p}{1 + 3 X_H + 4 X_H x_i},
\end{equation}
where $m_p$ is the proton mass and $k$ is the Boltzmann constant. We then computed the averages of the two quantities in radial bins $\hat b_j$:

\begin{equation}
n_{{\rm e},j} = V^{-1}(\hat b_j) \sum_{i \in \hat b_j} V_i n_{{\rm e},i}.
\end{equation}
%
\begin{equation}
T_{j} = m^{-1}(\hat b_j) \sum_{i \in \hat b_j} m_i T_{i}.
\end{equation}

The radial bins were chosen as 23 logarithmically spaced intervals between 0.001 and 10 $r_{200c}$, where $r_{200c}$ is the radius from the halo center encompassing 200 times the critical density of the Universe, $\rho_c$. As a derived quantity, we also computed the entropy as
\begin{equation}
S(\hat b_j) =  \frac{T(\hat b_j)}{n_{\rm e}(\hat b_j)^{2/3}} . 
\end{equation}

\subsection{Magneticum simulations}

The Magneticum simulations used in this work were performed with the TreePM/SPH code P-Gadget3, an extended version of P-Gadget2 \citep{Springel2005}. In addition to hydrodynamics and gravity, the simulations also account for a variety of physical baryonic processes, including radiative cooling and heating from a time-dependent UV background~\cite[][]{haardt2001}, star formation and feedback~\cite[][]{springel2003}, metal enrichment from stellar evolution~\cite[][]{tornatore2004,tornatore2007}, and black hole growth and gas accretion powering energy feedback from AGN~\cite[][]{springeldimatteo2005,dimatteo2005,fabjan2010}. In this study, we consider a sample of 22,254 galaxy groups and clusters with mass $M_{500c} > 5\times 10^{12} M_{\odot}$
identified using the SubFind algorithm~\cite[][]{springel2001,dolag2009} in the ``Box2\_hr'' simulation box. This covers a comoving volume of $(352 h^{-1} {\rm cMpc})^3$ and is resolved with $2\times1584^3$ particles (which correspond to mass resolutions of $m_{\rm DM} = 6.9 \times 10^8 h^{-1} M_{\odot}$ and $m_{\rm gas} = 1.4 \times 10^8 h^{-1} M_{\odot}$, for DM and gas respectively). A $\Lambda$CDM cosmology with $h = 0.704$, $\Omega_b = 0.0451$, $\Omega_m = 0.272$, $\Omega_\Lambda =0.728$ and $\sigma_8 = 0.809$ is employed~\cite[from the 7-year results of the Wilkinson Microwave Anisotropy Probe][]{komatsu2011}. See also~\cite{biffi2022} for further details.

We compute gas thermodynamical profiles by selecting the hot, diffuse gas component in each halo of the sample, characterized by temperature higher than $5 \times 10^5 K$ and non-star-forming, representing the X-ray emitting intra cluster and group mediums. The gas mass-weighted temperature and electron density radial profiles are then generated by considering three-dimensional radial bins. The profiles of massive halos are resolved with 50 linear radial bins up to $1.5\,r_{500c}$. For smaller systems resolved with fewer gas particles, we instead adopt an equal-particle binning of the radial profiles, to ensure statistically reliable estimates. In all cases, we ensure a minimum of 150 selected gas particles in each radial shell.

\subsection{OverWhelmingly Large Simulations}
The OverWhelmingly Large Simulations \citep{Schaye2010} used in this work is a cosmological hydrodynamical simulation that was performed using an extended version of the ThreePM/SPH code Gadget3 \citep{Springel2005}. The simulations include radiative cooling \citep{Wiersma2009a}, star formation \citep{Schaye2008}, stellar evolution and chemical enrichment \citep{Wiersma2009b}, kinetic supernovae feedback \citep{Dalla2008} and feedback from supermassive black holes \citep{springeldimatteo2005,Booth2009}. The simulations employ a flat $\Lambda$CDM cosmology with $h= 0.73$, $\Omega_{\rm b}=0.0418$, $\Omega_{\rm m}=0.238$, $\Omega_\Lambda=0.762$ and $\sigma_8=0.74$ that are taken from the analysis of 3-year results of WMAP \citep{Spergel2007}. The entropy and temperature measurements of galaxy groups in OWL simulations are taken from \citet{McCarthy2010} where $\sim200$ galaxy groups are selected by applying a mass criterion of $M_{200c}>10^{13}~{\rm M_{\odot}}$. The temperature and electron density profiles are obtained by emission-weighting the gas properties, and the entropy measurements are obtained by combining the temperature and electron density measurements.

They ran their simulations two times. In their first run (AGN run), they included all the sub-grid physics listed above, and in their second run (REF run), they turned off the AGN feedback. Both runs are indistinguishable in all aspects, with the only difference being that the former includes feedback from supermassive black holes as prescribed in \citep{Booth2009}. We refer the reader to \citet{Schaye2010} and \citet{McCarthy2010} for further details on the OWL simulations and the measurements of the thermodynamic properties of the galaxy groups in OWLS.

\subsection{Comparisons of observations with OWL, MillenniumTNG, and Magneticum simulations}

A fair comparison between simulations and observations is only possible when the compared samples are subject to a similar selection and the simulated temperatures are weighted with a well-calibrated temperature weighting scheme that would reproduce the observed spectroscopic-like temperatures. We achieve this by using the \rosi\ selection function that is obtained by applying the same cleaning procedure described in Sect.~\ref{sec:sample_selection} to the mock catalogs from the \rosi's digital twin \citep{Seppi2022}. The eRASS1 selection function, encapsulating the selection and cleaning information, provides detection probabilities of the group- and cluster-scale dark matter halos as a function of $M_{500c}$ and $z$ \citep[see][for details]{Clerc2024}. The selection function obtained through this procedure is applied to the MillenniumTNG and Magneticum clusters and groups, and the entropy profiles of the simulated samples are measured at the three characteristic radii. Furthermore, the spectroscopic-like characteristic temperatures, $T(r<r_{500c})$, are obtained by weighting and projecting the simulated 3D temperature profiles of MillenniumTNG and Magneticum halos, using the temperature weighting scheme calibrated for \rosi\ \citep{ZuHone2023}. Lastly, we compare the entropy and characteristic temperature measurements of simulated groups with the observations.

Before comparing our results with the simulations employing different AGN feedback implementations, we first test the reference (REF) run of the OWL simulations of \citet{McCarthy2010} with the observations. The predictions of the REF run serve as a baseline simulation and can be used to test the overall impact of the AGN feedback on the entropy profiles and characteristic temperatures of groups \citep{Schaye2010}. Their REF run is performed with their full model, identical to their final run, but does not include AGN feedback. To have a fair comparison, we convert the core-excised characteristic temperature measurements in simulations ($T(0.15r_{500c}<r<r_{500c})$) to the core-included temperatures, $T(r<r_{500c})$, following the same approach presented in Sect.~\ref{sec:results}. After the correction, we compare our measurements and the predictions of the REF run of the OWL simulations at the three characteristic radii, as shown in Fig.~\ref{fig:OWL_simulations_comp}. We find that the main visible effect of AGN feedback is to lift the entropy values of the groups outside the core (e.g., at $r_{2500c}$ or $r_{500c}$) for all groups, with a more marked effect at the lowest temperatures. The excess entropy induced by the AGN feedback in observations leads to significant disagreement between our measurements and the predictions of their REF run outside the core. On the other hand, the consistency with the AGN run (the twin simulations of the REF run, but the AGN feedback is turned on) suggests that the observations significantly favor the presence of strong AGN feedback in galaxy groups. It can further be seen from Fig.~\ref{fig:OWL_simulations_comp} that at the core $0.15r_{500c}$, the data points of \citet{McCarthy2010} for the AGN feedback run are indistinguishable from data points of the reference run, and their results agree well with ours at this radius. Contrarily, outside the core (at $r_{2500c}$ or $r_{500c}$), the reference and the feedback runs are significantly different at the low-temperature parameter space, indicating the strong AGN feedback imprint. This suggests that the entropy measurements at the core probe the AGN feedback much less efficiently than those outside the core at the group scale halos, which is also shown in \citet{McCarthy2010} for core-excised quantities. We note that the agreement at the core of groups between the reference and AGN feedback runs are obtained from the OWL simulations, where a thermal AGN feedback model is implemented. It is unclear whether such a conclusion would hold true for simulations implementing a feedback model that includes both kinetic and thermal feedback that leads to significant turbulent stirring in the core, such as MillenniumTNG. Therefore, to verify the universality of the conclusion, further tests are required in the future on simulations implementing kinetic and thermal feedback.

After comparing our measurements with the (REF) run of the OWL simulations, we compare our results with the three simulations employing different AGN feedback implementations: Magneticum, MillenniumTNG, and OWL (AGN run). The comparison between our measurements and the predictions of the simulations at the three characteristic radii is shown in Fig.~\ref{fig:Simcomps_T}.

At the cores of groups ($0.15r_{500c}$), we find the \rosi\ observations agreeing relatively well with the OWL simulations between the IGrM temperatures $T=0.73-1.79~\rm keV$ whereas at the warmer IGrM temperatures $T=1.79-2.68~\rm keV$ the observations fall above their predictions. We further find that the Magneticum and MillenniumTNG simulations overpredict the average entropy for the cool/lower-mass groups ($T=0.73-1.44~\rm keV$), whereas the agreement becomes better in the MillenniumTNG at the warmer IGrM temperatures, close to the cluster ICM temperature range. Even though our measurements disagree with the average entropy predictions of Magneticum and MillenniumTNG, our measurements still lie within the $1\sigma$ scatter of the simulated profiles.

At the mid-region of groups ($r_{2500c}$), the entropy measured by \rosi\ agrees well with the Magneticum and OWL simulation for cool/lower-mass groups, whereas the MillenniumTNG simulations underpredict the entropy for the groups in this region at a $\sim2.5\sigma$ level. We find that our measurements also agree well with the Magneticum simulations for warm/higher-mass groups. Furthermore, we find that as the temperatures/masses of the groups increase ($T>1.44$~keV) and approach the cluster temperatures/masses, the offset between observations and the predictions of MillenniumTNG and OWL simulations becomes more significant, starting from a statistical disagreement level of $3.5\sigma$ at $T=1.44~\rm keV$ and going up to a level of $8.5\sigma$ at $T=2.68~\rm keV$.

At the group outskirts ($r_{500c}$), we observe an overall agreement between the \rosi\ observations and the Magneticum simulations at all temperatures. While the entropy measurements of galaxy groups in MillenniumTNG simulations are slightly below observations, we find that they are consistent at a $2\sigma$ confidence level at all temperatures. Although our measurements are in $1\sigma$ agreement with the OWL simulations for cooler/lower-mass groups $T<1.44$~keV, the departure from observations becomes more significant ($>2\sigma$) at warmer temperatures, close to the ICM regime. 

The AGN feedback implementations in most large-scale simulations are variations of the ~\cite[][]{springeldimatteo2005,dimatteo2005} model. Even though similar in spirit, different adaptions and extensions to the original model have been made for the different simulations, leading to a strongly varying impact on the hot gas among the three simulations covered in this work. While the main feature behind the model for the OWL simulation is to achieve an effective AGN feedback by accumulating the injected energy until a $\Delta T_{heat}$ is reached, Magneticum and MillenniumTNG introduce a transition from quasar-mode to a stronger radio mode feedback \cite[e.g., see][]{Sijacki2007,fabjan2010} instead. However, while in Magneticum the radio mode feedback is still modeled as a isotropic, thermal feedback with an increased efficiency \citep[e.g., see][]{Hirschmann2014}, in MillenniumTNG it is modeled as a kinetic feedback \citep[e.g., see][]{Pillepich2018, Weinberger2017}. In addition to these differences, the detailed setting within the individual models cannot only significantly change the distribution and properties of gas within galaxies, but also impact the hot gas and especially the entropy within and around clusters and groups \citep[e.g., see figure 3 in][]{fabjan2010}. Subsequently, in an updated version of the OWL simulations \citep{LeBrun2014}, three different $\Delta T_{heat}$ parameters ($10^{8}$, $10^{8.5}$ and $10^{8.7}$~keV) of the \citet{Booth2009} model, coupled with the burstiness and the energeticness of the feedback, are tested. The authors report that $\Delta T_{heat}$ can be used to tune the impact of the feedback in the IGrM gas. They further show that for groups, the normalization and slope of the $S-T$ relation (or the $S-M$ relation) does not depend much on $\Delta T_{heat}$ at the core ($0.15r_{500c}$). In comparison, an increase in the $\Delta T_{heat}$ parameter corresponds to a larger normalization and possibly a different slope at the outskirts ($r_{2500c}$ and $r_{500c}$) due to the low-entropy gas getting ejected \citep[see Fig. 6 in][]{LeBrun2014}. This can also be seen in Fig.~7 of \citet{LeBrun2014} where the gas gets ejected in the case of more bursty and energetic feedback and results in lower densities at all radii out to $1.6r_{500c}$. 

In light of these indications found within simulations, the discrepancy between the observed entropy and the predictions of MillenniumTNG and OWL simulations at the outskirts can be justified by the effectiveness and energetics of the individual AGN feedback models being not sufficient to match our observations at the group scale. The weaker AGN feedback found in MillenniumTNG, leading to lower entropy in the outskirts, may also be related to the elimination of the magnetic fields from the TNG physics model, as noted in \citet{Pakmor2023}. The agreement between the observations and the Magneticum simulations in the group outskirts suggest that the underlying model is able to effectively treat a broad range in mass, extending the previously reported good agreement of entropy profiles for galaxy clusters \citep{Planelles2014,Rasia2015} to group scales. 

The agreement between the entropy measurements in \rosi\ observations and the numerical simulations at the cores is better than the outskirts. Slightly higher entropy in simulations is harder to explain using the \citet{Booth2009} model since the normalization and the shape at the core ($0.15r_{500c}$) do not seem to be affected by the change of the $\Delta T_{heat}$ parameter in the runs presented in \citet{LeBrun2014}. 

We note that the interpretations above for the OWL simulation assume that the \citet{Booth2009} model can reproduce the gas properties by tuning the $\Delta T_{heat}$ parameter relatively accurately, which may or may not be the case. Furthermore, as also discussed above, the AGN feedback implementation in MillenniumTNG and Magneticum simulations are even different (among them as well as compared to OWL). For these reasons, a more detailed analysis would be needed to better understand which aspects of the underlying AGN feedback models are driving the differences between the simulations and the \rosi\ observations, which is beyond the scope of this paper. Furthermore, we note that the selection function we used in this work is a function of mass and redshift, and the use of it significantly improves the robustness of our comparison. In future studies, a profile-dependent selection function can be used that would make the selection procedure applied to the simulations even more realistic. Ultimately, the most accurate comparison would be achieved by producing synthetic eROSITA observations through fully forward modeling cosmological hydrodynamic simulations to which the same detection and data analysis pipeline used for real observations can be applied.

\section{Conclusions}
\label{sec:conclusions}

Our work places the tightest constraints on the impact of AGN feedback on the average thermodynamic properties by populating the low-mass galaxy groups down to cool IGrM temperatures of 0.7~keV. \rosi's superb sensitivity in the soft X-ray band led to the detection of a large number of galaxy groups with a well-understood selection function. When stacked, the \rosi\ data provides unprecedented statistical power for the measurements of X-ray properties of these sources. We used a sample of 1178 galaxy groups to place constraints on the impact of AGN feedback on the thermodynamic properties of the IGrM. We selected the galaxy group sample based on the primary sample of the eRASS1 clusters and groups and applied a rigorous selection and cleaning. The cleaning procedure was designed to provide a pure sample with a well-defined selection function while maximizing the sample size. A Bayesian imaging analysis was carried out for all the 1178 galaxy groups in the sample, where the nearby clusters and bright point sources were co-fit. The galaxy groups with similar statistical and physical properties, such as count and temperature, were then grouped together into 271 bins. A joint Bayesian spectral fitting was performed on the groups in the same bin to increase the statistical power in each bin with the sources with similar properties. 

Constraining baryonic physics at group scales is a highly challenging task. Systematic effects must be considered to achieve a reliable and robust conclusion. We have quantified and discussed three major systematics for thermodynamic profile studies and conservatively took them into account for our final measurements and conclusions. Apart from the robust thermodynamic property measurements, we have also provided a detailed comparison of our findings with the state-of-the-art simulations by accounting for the selection effects. We assessed the agreement between our measurements and the simulations, employing various AGN feedback implementations to pave the way for more realistic AGN feedback modeling in numerical simulations. 

This paper is the first in-depth study of galaxy groups with \rosi\ focusing on the impact of AGN feedback on the entropy and temperature measurements within an overdensity radius. The main conclusions of this work are as follows:

\begin{enumerate}
\item[$\boldsymbol{-}$] With a sample of 1178 galaxy groups and average thermodynamic properties of 271 binned groups, our work stands as the most comprehensive study of the hot gas in galaxy groups in terms of sample size, diversity, and statistics. The selection effects have been considered for the first time while comparing the measured thermodynamic profiles of galaxy groups with the numerical simulations.
\item[$\boldsymbol{-}$] Overall, the entropy measurements at three characteristic radii, $0.15r_{500c}$, $r_{2500c}$, and $r_{500c}$, and the characteristic temperature of the galaxy groups detected in the \rosi\ first All-Sky Survey observations are in good agreement with the previously reported results in S09 and \citet{Johnson2009}, within the uncertainties (see Sect.~\ref{sec:systematics}). The largest mismatch between the \rosi\ measurements presented in this work and the \chandra measurements presented in S09 is at the low-temperature parameter space ($T(r<r_{500c})<1.15$~keV). We argue that this is because the completeness and the selection of the two samples is different in this temperature range. The archival sample used in S09 consists mostly of bright and morphologically undisturbed galaxy groups detected in \rosat observations and was followed up by \chandra. In contrast, the group sample used in this work has a more uniform and well-defined selection.
\item[$\boldsymbol{-}$] We compared our entropy measurements with the reference (REF) run of OWL simulations that include various non-gravitational processes but not the AGN feedback. From this comparison, we conclude that the impact of AGN feedback on the entropy profiles of groups is significant at the outskirts ($r_{2500c}$ and $r_{500c}$) and less pronounced near the core ($0.15r_{500c}$). This result suggests that for groups, AGN feedback has a larger impact on the outskirts than the core, contrary to its impact on the observed gas properties of clusters of galaxies \citep{LeBrun2014}. Due to their shallower potential wells, the feedback from the central black hole is able to move the low-entropy gas to much larger radii (e.g., to the outskirts) of galaxy groups. 
\item[$\boldsymbol{-}$] Our measurements have significant constraining power on the impact of AGN feedback on the thermodynamic properties of the IGrM gas and serve as a reference for the feedback implementations in numerical simulations and theoretical models. We compared our results with three state-of-the-art cosmological hydrodynamic simulations, including MillenniumTNG, Magneticum, and OWL, employing various AGN feedback implementations by accounting for the selection effects. In the cores of the galaxy groups, the entropy agrees well with OWL simulations out to $T=1.79~\rm keV$. Although the sample-averaged entropy from the MillenniumTNG and Magneticum simulations are higher, the measurements are within the scatter of the respective simulations. At the outskirts, ($r_{2500c}$ and $r_{500c}$), the observed entropy agrees well with the Magneticum simulations for groups with IGrM temperatures down to 0.79~keV. In OWL simulations, although we observed a similar entropy-flattening trend for the cooler groups, the departure from observations becomes significant toward the galaxy cluster regime at higher temperatures. A similar trend has been observed in MillenniumTNG simulations; however, in this case, the entropy offset is relatively significant for all temperature regimes. Overall, the AGN feedback implementation in Magneticum simulations reproduces our observations the best at the three characteristic radii. 
\end{enumerate}

This work demonstrates the potential of \rosi\ for exploring the baryonic physics at the galaxy groups out to large radii. Deeper data with the \rosi\ All-sky Survey will allow for the detection of a larger sample of galaxy groups pushing down in mass and IGrM temperature floors. Employing these groups in similar future studies with larger statistical power will enable the testing of hydrodynamical simulations in very cool temperatures below $<0.5$~keV, which are currently unreachable with the eRASS1 sample.

\begin{acknowledgement}
This work is based on data from \rosi, the soft X-ray instrument aboard SRG, a joint Russian-German science mission supported by the Russian Space Agency (Roskosmos), in the interests of the Russian Academy of Sciences represented by its Space Research Institute (IKI), and the Deutsches Zentrum f{\"{u}}r Luft und Raumfahrt (DLR). The SRG spacecraft was built by Lavochkin Association (NPOL) and its subcontractors and is operated by NPOL with support from the Max Planck Institute for Extraterrestrial Physics (MPE).

\\
The development and construction of the \rosi\ X-ray instrument were led by MPE, with contributions from the Dr. Karl Remeis Observatory Bamberg \& ECAP (FAU Erlangen-Nuernberg), the University of Hamburg Observatory, the Leibniz Institute for Astrophysics Potsdam (AIP), and the Institute for Astronomy and Astrophysics of the University of T{\"{u}}bingen, with the support of DLR and the Max Planck Society. The Argelander Institute for Astronomy of the University of Bonn and the Ludwig Maximilians Universit{\"{a}}t Munich also participated in the science preparation for \rosi.

\\

The eROSITA data shown here were processed using the eSASS/NRTA software system developed by the German eROSITA consortium.

\\
E. Bulbul, A. Liu, V. Ghirardini, C. Garrel, S. Zelmer, and X. Zhang acknowledge financial support from the European Research Council (ERC) Consolidator Grant under the European Union’s Horizon 2020 research and innovation program (grant agreement CoG DarkQuest No 101002585). N. Clerc was financially supported by CNES. S. Bose is supported by the UK Research and Innovation (UKRI) Future Leaders Fellowship [grant number MR/V023381/1]. C. Hern\'andez-Aguayo acknowledges support from the Excellence Cluster ORIGINS which is funded by the Deutsche Forschungsgemeinschaft (DFG, German Research Foundation) under Germany's Excellence Strategy -- EXC-2094 -- 390783311.
\\
This work made use of the following Python software packages: 
Astropy\footnote{https://www.astropy.org/} \citep{Astropy2022}, 
Colossus\footnote{https://bdiemer.bitbucket.io/colossus/} \citep{Diemer2018},
emcee\footnote{https://emcee.readthedocs.io/} \citep{emcee},
Matplotlib\footnote{https://matplotlib.org/} \citep{Hunter2007matplotlib}, 
NumPy\footnote{https://numpy.org/} \citep{Harris2020numpy},
pyfftw\footnote{https://pypi.org/project/pyFFTW/} \citep{FFTW05}
PyXspec\footnote{https://heasarc.gsfc.nasa.gov/docs/xanadu/xspec/python/html/} \citep{Arnaud1996},
SciPy\footnote{https://scipy.org/} \citep{Virtanen2020SciPy}, 
Vorbin\footnote{https://pypi.org/project/vorbin/} \citep{Cappellari2003},
\end{acknowledgement}

%
%

\bibliographystyle{aa}
\bibliography{groups}

\begin{appendix} 

\section{Energy band selection for the imaging analysis}
\label{sec:ebandselection}

We chose the energy band for the imaging analysis by optimizing the signal-to-noise ratio given multiple X-ray foreground and background components. The models we used during this optimization scheme include 1) a 0.1~keV unabsorbed {\sc APEC} component for the Local Hot Bubble, whose flux is scaled to $2.3\times10^{-13}$ erg s$^{-1}$ cm$^{-2}$ deg$^{-2}$ in the 0.3--0.7~keV band \citep{Yeung2023}; 2) a 0.18~keV absorbed {\sc APEC} component for the Galactic Halo, whose absorbed flux is scaled to $1.1\times10^{-12}$ erg s$^{-1}$ cm$^{-2}$ deg$^{-2}$ in the 0.5--2.0~keV band \citep{Henley2013}; 3) a $\Gamma=1.4$ power law component for the Cosmic X-ray Background component, whose absorbed flux is scaled to $6\times10^{-13}$ erg s$^{-1}$ cm$^{-2}$ deg$^{-2}$ in the 0.5--2.0~keV band, which corresponds to a $10^{-13}$ erg s$^{-1}$ cm$^{-2}$ point source flux cut; and 4) the instrumental background from eROSITA filter-wheel-closed observations. The foreground HI column density is fixed to $2.7\times10^{20}$ cm$^{-2}$ for this study, which is the averaged value of the sample. With these foreground and background configuration, we find that the signal-to-noise ratio of a $T=1$~keV {\sc APEC} source component at the redshift of 0.18 reaches the maximum in the 0.3--1.8~keV band at a source-to-background ratio of 1. We note that the upper boundary of the optimal band could increase if we adopt a higher source-to-background ratio since the $\log N - \log S$ curve of the selected sample is a power low. Nevertheless, we select a source-to-background ratio of 1 and we adopt the 0.3--1.8~keV band for our imaging analysis to maintain high signal-to-noise for the faint sources.

\end{appendix}

\end{document}